\documentclass[journal,transmag]{IEEEtran}

\usepackage{cite}
\usepackage{amsmath,amssymb,amsfonts}
\usepackage{algorithmic}
\usepackage{graphicx}
\usepackage{textcomp}
\usepackage{xcolor}
\def\BibTeX{{\rm B\kern-.05em{\sc i\kern-.025em b}\kern-.08em
    T\kern-.1667em\lower.7ex\hbox{E}\kern-.125emX}}

\usepackage{tikz}
\usetikzlibrary{patterns}
\usepackage{nicefrac}
\usepackage{prettyref}
\usepackage[capitalise]{cleveref}
\usepackage{mathtools}
\usepackage{multirow}
\usepackage[labelformat=simple]{subcaption}
\usepackage{booktabs}
\usepackage{diagbox}
\usepackage[linesnumbered,ruled]{algorithm2e}
\usepackage{csquotes}
\usepackage{array}
\usepackage{balance}

\graphicspath{{images/}}
\DeclarePairedDelimiter\ceil{\lceil}{\rceil}

\newrefformat{sec}{Section~\ref{#1}}
\newrefformat{fig}{Fig.~\ref{#1}}
\newrefformat{tab}{Table~\ref{#1}}
\newrefformat{lst}{Listing~\ref{#1}}
\newrefformat{alg}{Algorithm~\ref{#1}}

\newcolumntype{L}[1]{>{\raggedright\arraybackslash}p{#1}}
\newcolumntype{C}[1]{>{\centering\arraybackslash}p{#1}}
\newcolumntype{R}[1]{>{\raggedleft\arraybackslash}p{#1}}
\newcolumntype{$}{>{\global\let\currentrowstyle\relax}}
\newcolumntype{^}{>{\currentrowstyle}}
\newcommand{\rowstyle}[1]{\gdef\currentrowstyle{#1}%
	#1\ignorespaces
}

\begin{document}


\title{Enabling Seamless Device Association with DevLoc\\using 
Light Bulb Networks for Indoor IoT Environments}

\author{
\IEEEauthorblockN{Michael Haus}
\IEEEauthorblockA{
	\textit{Technical University of Munich}\\
	haus@in.tum.de}
\and
\IEEEauthorblockN{Jörg Ott}
\IEEEauthorblockA{
	\textit{Technical University of Munich}\\
	ott@in.tum.de}
\and
\IEEEauthorblockN{Aaron Yi Ding}
\IEEEauthorblockA{
	\textit{Delft University of Technology}\\
	aaron.ding@tudelft.nl}
}


\maketitle

\begin{abstract}
To enable serendipitous interaction for indoor IoT environments, 
spontaneous device associations are of particular interest so 
that users set up a connection in an ad-hoc manner. Based on the 
similarity of light signals, our system named DevLoc takes 
advantage of ubiquitous light sources around us to perform 
continuous and seamless device grouping. We provide a 
configuration framework to control the spatial granularity of 
user's proximity by managing the lighting infrastructure through 
customized visible light communication. To realize either 
proximity-based or location-based services, we support two modes 
of device associations between different entities: 
device-to-device and device-to-area. Regarding the best 
performing method for device grouping, machine learning-based 
signal similarity performs in general best compared to 
distance and correlation metrics. Furthermore, we analyze 
patterns of device associations to improve the data privacy by 
recognizing semantic device groups, such as personal and 
stranger's devices, allowing automated data sharing policies.
\end{abstract}


\section{Introduction}
The capabilities of wireless devices, such as laptops, mobile 
phones, tablets, IoT boards, enable flexible formation of ad-hoc 
groups. New opportunities arise for users in physical proximity 
by dynamic group association to spontaneously share resources or 
information. We support two different types of proximity 
applications aimed for end users and Internet of Things (IoT). 
We envision two use cases for user-oriented, proximity-based 
applications \cite{Narayanan.2011}: 1) Alice is a tourist, rides 
on the subway and wants to ask locals for the best way to the 
museum, and 2) Carol is a manager who wants to automatically 
record who is present at her daily meetings. In addition, we 
emphasize two use cases for proximity-based IoT applications: 1) 
location-tagged data from IoT boards facilitate data merging and 
filtering in case we have the same information from multiple IoT 
boards placed in the same area and 2) location-based access 
policy for consumer smart home platforms \cite{Mare.2019}, e.g., 
Amazon Echo or Google Home. To realize such applications, we 
explore proximity as a group association technique where devices 
find one another when they are brought within a close distance 
of each other or in a dedicated space \cite{Chong.2013}.
Thereby, proximity identifies potential group members and device 
association refers to the technique that connect (a subset of) 
those potential group members.

Our system for continuous and seamless device grouping named 
DevLoc uses visible light signaling because light sources are 
ubiquitous around us ensuring practicality. DevLoc combines 
visible light and Wi-Fi as the primary communication means. 
Compared to the electromagnetic waves of Wi-Fi which easily 
penetrate physical barriers, visible light does not pass through 
opaque objects and hence it is a good candidate to realize 
distance-bounding wireless communication.
Based on the distance-limited nature of visible light, we 
achieve more fine-granular device associations which are 
impossible to recognize with propagating Wi-Fi. To compensate 
the downsides of visible light communication (VLC), such as lack 
of hardware support at mobile devices, e.g., to receive the 
data, we provide the design of a light tag for pervasive VLC. 
The light tag usable as sticker can be easily attached to 
different end-user devices enabling light transmissions. 
Furthermore, by the analysis of log files of device associations 
we can infer different semantic device groups, e.g., personal 
devices, based on the frequency and time of device encounters. 
This allows us to automatically generate meaningful data sharing 
policies between devices associated with a certain type such as 
personal, family, etc. to define with whom sharing or 
aggregating data. Hence, we are able to move the task to specify 
data sharing policies to lower communication layers, usually 
handled as part of the application layer in wireless systems 
used today.
Moreover, to minimize the adaption effort to introduce DevLoc, 
we adopt a master-slave principle for light bulbs of existing 
lighting infrastructure. Only the master light bulb requires 
computation power to perform device associations, the slave 
light bulb simply broadcasts the light pattern received from the 
master light bulb via a Wi-Fi interface.

In contrast to existing systems for device grouping, we provide 
a complete framework to manage the lighting infrastructure and 
control the spatial granularity of device grouping. As a result, 
we are able to facilitate applications with different spatial 
expansion of proximity and overcome the main disadvantage of 
location tags \cite{Narayanan.2011} that users have no control 
over the spatial granularity of proximity. We enrich the 
lighting infrastructure by adding light signaling to the widely 
used Light-Emitting Diode (LED) lamps in residential and office 
settings. To automatically link physically nearby devices based 
on the similarity of light patterns, our custom light bulbs 
combine illumination with visible light signaling.
Our generated light patterns for device grouping are 
unpredictable nonces associated with a location 
\cite{Narayanan.2011}. For instance, like a shared pool of 
entropy between all users at a given location at a given time. 
In the following, we state the two key properties of light 
patterns for device grouping \cite{Narayanan.2011}: 1) 
reproducibility meaning that two measurements at the same place 
and time match with high probability, and 2) unpredictability so 
that an adversary at another location is unable to produce a 
location tag that matches the tag measured at the actual 
location at that time.

In a nutshell, our work makes the following contributions:
\begin{enumerate}
	\item We introduce DevLoc for seamless device grouping 
	taking advantage of boundary-limited visible light 
	signaling. We build a custom light bulb to enrich the 
	lighting infrastructure being able to control the spatial 
	granularity of user's proximity.
	\item To qualify the feasibility of real-world deployments 
	of DevLoc, we analyze the propagation characteristics of 
	VLC, e.g., maximum achievable detection range of light 
	patterns. Furthermore, we perform a feature selection for 
	light patterns and we analyze the performance of several 
	signal comparison methods via two simulations for static 
	device-to-device grouping and dynamic device-to-area 
	grouping.
	\item To enhance data privacy and ease the setup of data 
	sharing, we extract different features from logs of device 
	associations and classify them to infer semantic device 
	groups such as personal and stranger's devices. On this 
	basis, we are able to create data sharing policies like 
	sensitive information can be only shared among personal 
	devices.
\end{enumerate}

\section{Related Work}
DevLoc deals with the areas of device coupling, device grouping, 
device association, and device pairing. Our device association 
is a guidance technique without human interaction on the basis 
of user's proximity in the real world. The work of 
\cite{Chong.2014} provides an overview by classifying techniques 
for device grouping in the following way: 1) input aims at user 
actions like triggering commands, entering data, or direct by 
manipulating, 2) enrollment uses on-time registration of devices 
with an identity, 3) guidance takes advantage of users acting in 
the real world to link devices via contact or alignment, and 4) 
matching involves different approaches where users compare the 
output of the involved devices to acknowledge a connection.

Visible light positioning such as in \cite{Kuo.2014, Yang.2015b, 
Yang.2015c} is out of scope because we are not interested in the 
user's position to protect the user's privacy. We only need to 
infer whether users are nearby. Hence, we use context 
information such as ambient light to recognize proximate devices 
based on the distance-limited nature of light. In this 
regard, to detect the co-presence of devices, other system 
approaches use ambient audio \cite{Schurmann.2013}, ambient 
noise and luminosity \cite{Miettinen.2014}, accelerometer data 
caused by hand shaking \cite{Mayrhofer.2009}, radio signals 
\cite{Varshavsky.2007}, magnetometer readings of very close 
devices \cite{Jin.2016}, and gait cycle detection of moving 
users \cite{Schurmann.2018}. The existing work aims to link 
mainly two devices whereas DevLoc enables group associations. 
Group association is not simply an extension of pairwise 
association with additional devices \cite{Chong.2013}. Rather 
than multiple device pairings, many people expect that group 
association is a single-step procedure. The user study of 
\cite{Chong.2013} states that groupwise associations are not 
rated highly for simplicity, but close proximity is a popular 
technique to link devices.



\section{Device Associations via DevLoc System}

\begin{figure}
	\centering
	\begin{subfigure}[b]{\linewidth}
		\centering
		\includegraphics[width=0.93\linewidth]{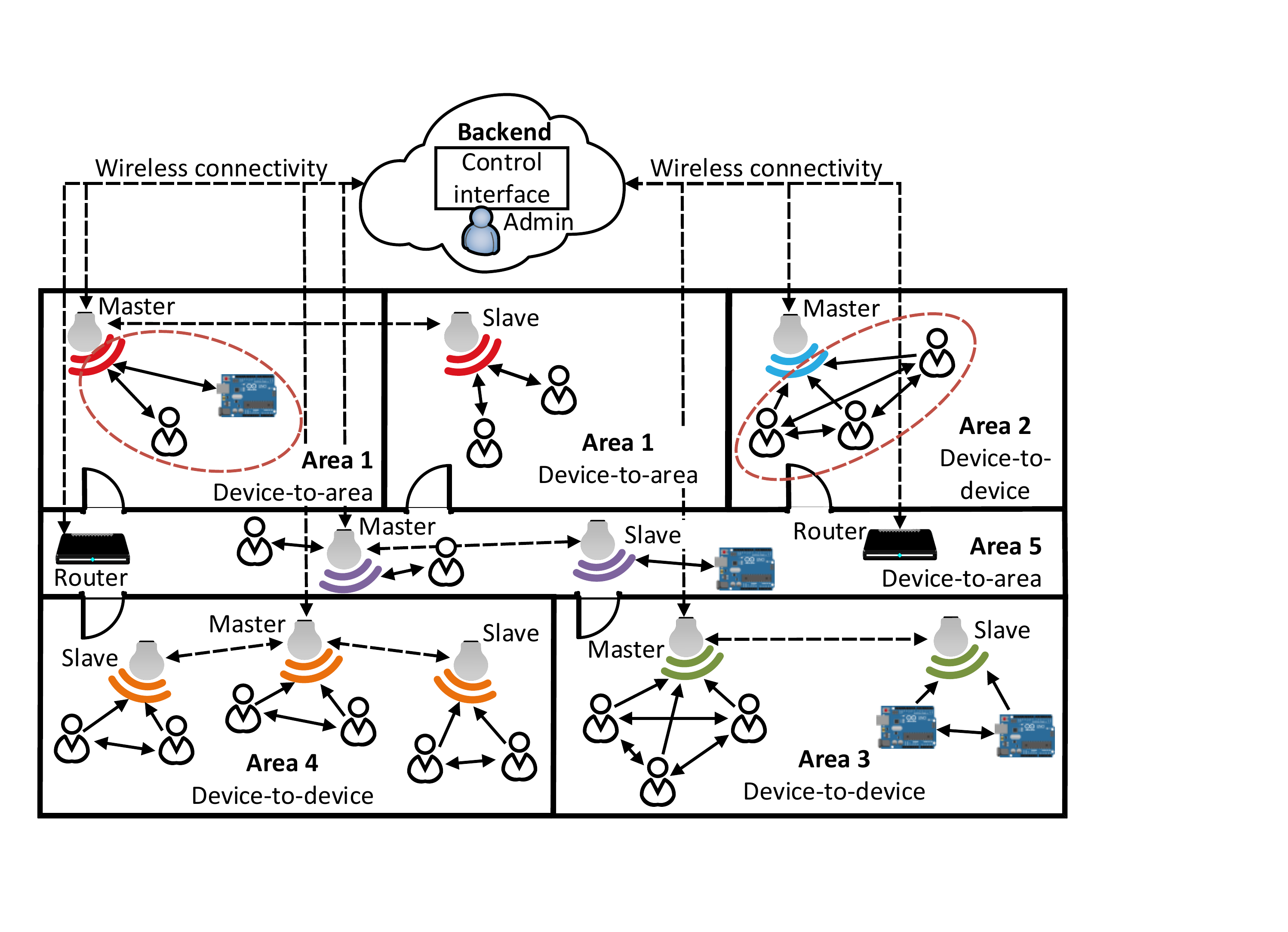}
		\caption{DevLoc's system combines Wi-Fi routers and 
		light bulbs to enable device associations for mobile 
		users and IoT boards.}
		\label{fig:system-device-grouping}
	\end{subfigure}
	\\
	\vspace{2mm}
	\begin{subfigure}[b]{\linewidth}
		\centering
		\includegraphics[width=0.318\linewidth]{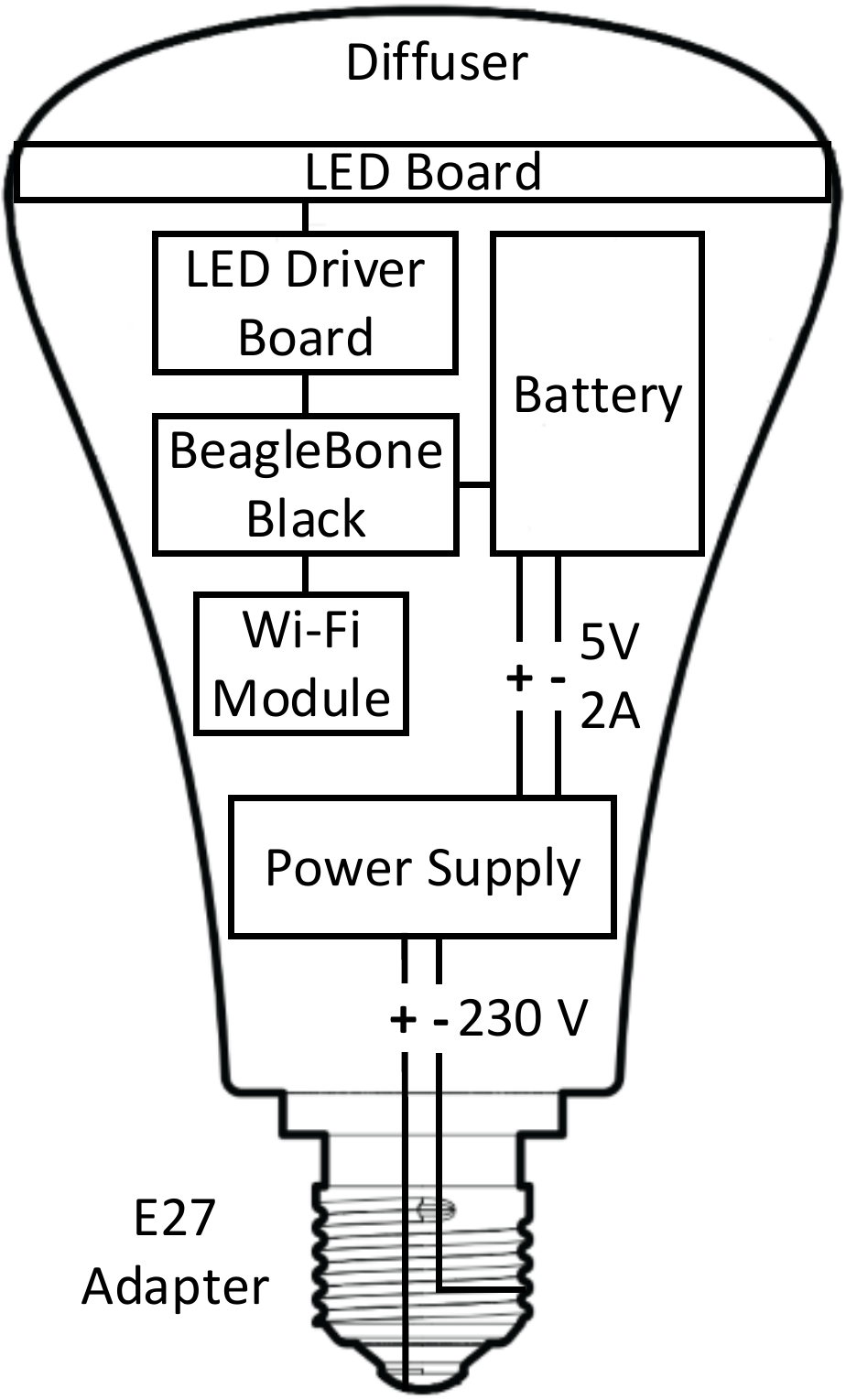}
		\qquad
		\includegraphics[width=0.24\linewidth]{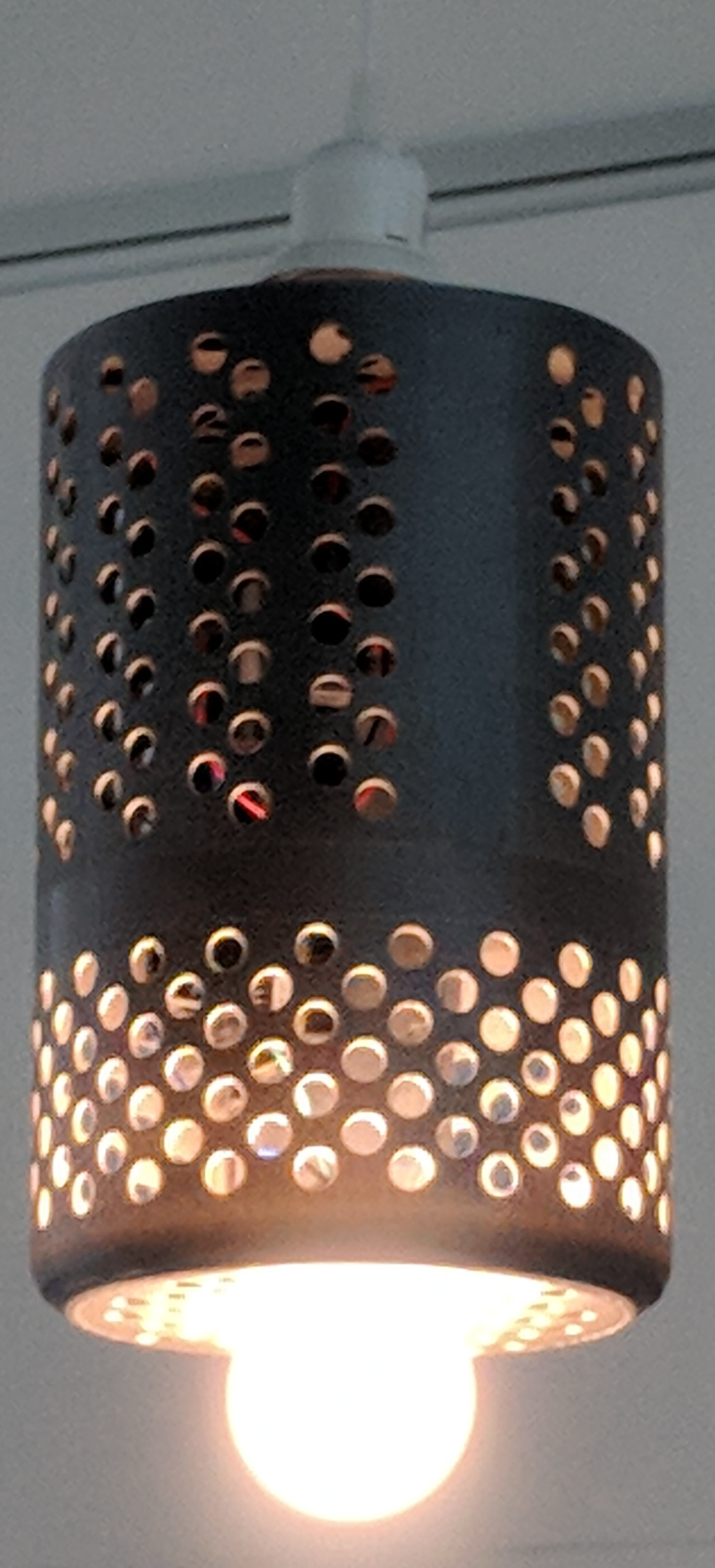}
		\caption{DevLoc is based on the customized LocalVLC 
		system \cite{Haus.2019b} to realize visible light 
		signaling for device grouping. We present the hardware 
		platform of the deployed 3D-printed light bulb.}
		\label{fig:custom-light-bulb}
	\end{subfigure}
	\caption{Enable seamless device associations using DevLoc}
\end{figure}

\prettyref{fig:system-device-grouping} presents the DevLoc 
framework for device grouping to combine radio-based 
communication like Wi-Fi covering larger areas and non 
radio-based communication such as light which is spatially more 
fine-grained due to walls, doors. We enrich existing lighting 
with visible light signaling for device grouping. Based on a 
master-slave principle of the light bulbs, we are able to 
semantically link multiple rooms or regions and thereby flexibly 
control the user's spatial granularity. We use different colors 
in \prettyref{fig:system-device-grouping} to illustrate varying 
light patterns at the light bulbs for device associations. The 
dotted red circles highlight the association among different 
entities: a) device-to-device using only the device's light 
signals or b) device-to-area using the device's light signal and 
an area's reference light signal for signal comparison. The goal 
of DevLoc is to ease data sharing among mobile 
user devices like tablets, smartphones, laptops, and static IoT 
boards. Inspired by \cite{Schmid.2014b, Gummeson.2017, 
Schmid.2015} and as central part of DevLoc, our custom light 
bulb in \prettyref{fig:custom-light-bulb} establishes a Wi-Fi 
link to the lighting configuration framework and broadcasts 
light patterns at a high frequency. On this basis, we can 
replace existing illumination units and hence we are able to 
restrict the problem of light pollution, where different visible 
lights would be overlapping for illumination and communication.
We now describe in more detail the setup and working principle 
of DevLoc.

\subsection{Adaptable Spatial Granularity of Device Grouping}
DevLoc allows to select proximity areas to define the geographic 
structure of the device associations. For example, 
\prettyref{fig:system-device-grouping} uses room numbers for 
area one and region names like corridor for area five. The 
lighting configuration framework runs at the backend and, 
initially, each light bulb and Wi-Fi router registers itself 
with the backend. As a result, DevLoc knows all light bulbs and 
their specific areas and randomly selects for each region one of 
the light bulbs as master light bulb, the remaining ones act as 
slaves. The backend randomly creates a light pattern for each 
registered master light bulb and the slave(s) broadcast the same 
light pattern with the master-slave mechanism for the light 
bulbs. We can dynamically choose the spatial granularity of 
device proximity by adapting the groups of light bulbs covering 
different regions. We can use the same light pattern over 
different rooms which are semantically the same region, e.g., 
area one in \prettyref{fig:system-device-grouping} to link two 
rooms. The size of rooms and regions like corridors, and the 
number and distribution of light bulbs define the achievable 
spatial granularity of device groupings. To achieve the most 
fine-granular user proximity, each light bulb works on its own 
as master. In our experiments, we identified the communication 
range of our custom light bulb of up to 10\,m. Moreover, the 
master-slave mechanism of our light bulbs allows a minimum of 
technical adaptions on existing illumination. Only the master 
light bulbs need computing power to perform the device 
groupings, the slave light bulbs require only a radio 
connection, e.g., Wi-Fi or Bluetooth, to receive the commands 
from the corresponding master bulb.

\subsection{Triggering Device Grouping}
We combine each master light bulb with a Wi-Fi router as a 
channel to the central configuration framework to maintain light 
patterns and for later device interaction. The light bulb 
continuously monitors the wireless connections of 
the Wi-Fi router and triggers device groupings. Due to the 
larger Wi-Fi coverage, one router can be combined with multiple 
master light bulbs. If there are no device groups yet and the 
Wi-Fi connections are changing, each linked master light bulb 
requests the continuously broadcasted light pattern received 
from the client(s). After receiving the client's data, the 
master light bulb initiates the device grouping to infer which 
devices are in 
the same light communication range instead of being only in the 
same Wi-Fi coverage. In case of a new Wi-Fi client, the master 
light bulb runs the signal matching to infer the matching device 
group without affecting other devices. When a Wi-Fi device 
disappears at the router, the master light bulb deletes this 
single client from existing device groups.

Moreover, user mobility may also trigger device groupings. For 
static users who don't move between rooms it is enough to 
observe the Wi-Fi connections for device grouping. In contrast, 
we need to manually start the device association via a 
predefined period, e.g., every few seconds, if users move 
between multiple regions but still connected to the same Wi-Fi 
router. To update the device grouping, we do not use signal 
strength changes of the user's Wi-Fi connection because it can 
change unexpectedly and yields excessive false positives and 
false negatives causing frequent device grouping updates.


\subsection{Two Modes for Device Grouping: Device and Area}
To support either location-based services (LBS) or 
proximity-based services (PBS), we are able to specify the mode 
of device groupings for each master light bulb: device-to-area 
grouping for LBS and device-to-device grouping for PBS. LBS 
needs to answer the question \enquote{where we are?} 
based on the absolute position of a user. In contrast, PBS needs 
to answer the question \enquote{who are we with?} based on 
context information to find co-location with other points of 
interest.
We encounter three main differences between 
device-to-device and device-to-area groupings: 1) trigger point 
in time of the device association, 2) required number of user 
clients for device association, and 3) signal comparison between 
different entities which affect the resulting binding either 
device-to-device or device-to-area.

For device-to-device groupings we need at least two connected 
user clients at the Wi-Fi router to start the device grouping. 
To link a Wi-Fi client to a specific device group, the master 
light bulb randomly selects one client from each existing device 
group for signal matching. The participating user clients only 
know which other clients are nearby and not at which indoor 
region they are located. Thereby, we can only realize PBS like 
data sharing among close-by users and LBS are not feasible, 
e.g., sharing the menu of the cafeteria since the users are 
nearby to the canteen, because location-related information is 
missing using the device-to-device grouping.

For device-to-area groupings, after the user client connected to 
the Wi-Fi router the corresponding master light bulb(s) 
immediately start the device association and compare the 
client's signal to the area's reference signal. We achieve a 
direct binding between the device and area. Thereby, we know 
which device is in which area and at the same time which other 
devices are close-by. There is no limitation with respect to the 
number of connected user clients, e.g., at least two connected 
clients for device-to-device association. In general, 
device-to-device associations provide less location-specific 
information compared to device-to-area groupings.

\subsection{Generation and Recognition of Light Patterns}
Our custom light bulb emits randomly generated light patterns 
for device associations. We independently create a random series 
of light on and off periods and combine them resulting in a 
light pattern. The duration of each light on and off period is 
in the range of [1, 5]\,ms. The minimum duration is constrained 
via the hardware of our light receiver, specifically, how fast 
the photodiode can be sampled. The maximum duration of each 
light on and off period 
is determined by avoiding unpleasant visual experience where 
light flickering effects are visible by human eyes. The light 
sender emits the light pattern in a loop for a restricted amount 
of time. To be able to differentiate light patterns, the length 
of the light pattern must be a multiple of two, i.e., after each 
light on period appears a light off period. To enhance the 
recognition rate of light patterns at the light receiver, we 
introduce a 10\,\% duration difference among light on and off 
series so that the time periods are sufficiently distinct. The 
photodiode at the light receiver samples the raw light signal as 
voltage in mV: a higher voltage refers to a light on period and 
a lower voltage refers to a light off period.


\textbf{How to detect reoccurring patterns in the light signal?} 
We use the cycle detection algorithm from \cite{Schurmann.2018} 
to find repeating patterns in our light signal. The algorithm 
supports signal matching of arbitrary co-aligned sensor data and 
reaches a reliable signal segmentation based on normalization. 
The algorithm's input expects a vector of voltage amplitudes $z 
= (z_1, ..., z_n)$ and the result is a sequence of consecutive 
light signal patterns. We use auto-correlation and distance 
calculation to find repeating signal parts. We can efficiently 
calculate the auto-correlation via the Wiener-Khinchin theorem 
\cite{Cohen.1998} with complexity $n \cdot log(n)$
\begin{align*}
F_R(f) &= FFT[z] \\
S(f) &= F_R(f) \cdot F_R^*(f) \hspace{5mm}
*\hat{=}\,\text{conjugate} \\
R(\tau) &= IFFT[S(f)]
\end{align*}
where $z$ are the voltage amplitudes. The auto-correlation 
$R(\tau)$ results in $m$ non-ambiguous local maxima $\zeta = 
\text{arg max}(R(\tau)) = \{\zeta_1, ..., \zeta_i ..., 
\zeta_m\}$. We compute the distances among all local maxima and 
a mean distance
\begin{align*}
\delta_{\text{mean}} 
= \ceil*{\frac{\sum_{i=1}^{m-1}\zeta_{i+1}-\zeta_i}{m-1}}
\end{align*}
at which $\delta_{\text{mean}}$ can be used to choose minima 
indices from $z$ representing signal patterns. To find the local 
minima $\mu$, every local maximum specifies a start point and 
$\delta_{\text{mean}}$ a search range
\begin{align*}
\mu &= \{\mu_1, ... \mu_i, ..., \mu_{m-1}\} \\
\mu_i &= \text{arg min}(z_{\zeta_i}, z_{\zeta_i+1}, ..., 
z_{\zeta_i+\delta_{\text{mean}}})
\end{align*}
Each $\mu_j$ refers to an index of a minimum in $z$ restricted 
to the range of $\delta_{\text{mean}}$. The indices in $\mu$ are 
used to separate the voltage amplitude $z$ into light cycles
\begin{align*}
Z &= \{Z_1, ..., Z_i, ..., Z_q\}  \\
Z_i &= (z_{\mu_{\frac{i}{2}}}, ..., z_{\mu_i}, ..., 
z_{\mu_{\frac{i+1}{2}}-1}) \text{ with } i = \{2, 4, ..., q\}
\end{align*}
In our experiments, the rate of successfully extracted light 
patterns decreases significantly in case of sudden changes of 
light patterns caused by light interference.
Therefore, we implement our own method to recognize light cycles 
considering the period of each light on and off phase. We define 
the light signal as a list of periods
\begin{align*}
\hat{z} = \{ (s_1, d_1), ..., (s_n, d_n) \}
\end{align*}
where $s_n \in \{0, 1\}$ specifies if the light is on or off and 
$d_i \in \mathbb{Z}$ refers to the duration of each period. We 
combine similar signal parts with a difference smaller than 
10\,\% because the light sender introduced a 10\,\% signal 
margin between the light on and off periods to improve 
the robustness of signal pattern detection. The remaining unique 
signal parts define the signal pattern including the period of 
each phase.
To identify the light pattern, we overlay the light signal with 
a time window specified by the pattern length which is defined 
for the system.

\begin{figure}
	\centering
	\begin{subfigure}[b]{0.49\linewidth}
		\begin{subfigure}[b]{\linewidth}
			\centering	
			\includegraphics[width=0.7\linewidth]{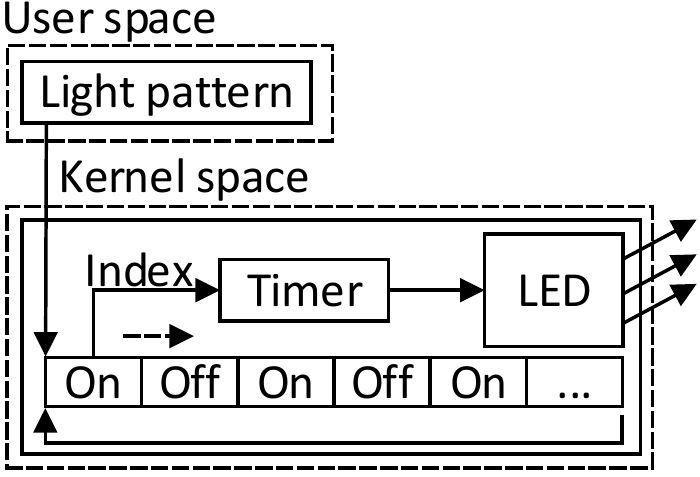}
			\caption{Scheme of light sender}
			\label{fig:light-sender-scheme}
		\end{subfigure}
		\\
		\\
		\begin{subfigure}[b]{\linewidth}
			\includegraphics[width=\linewidth]{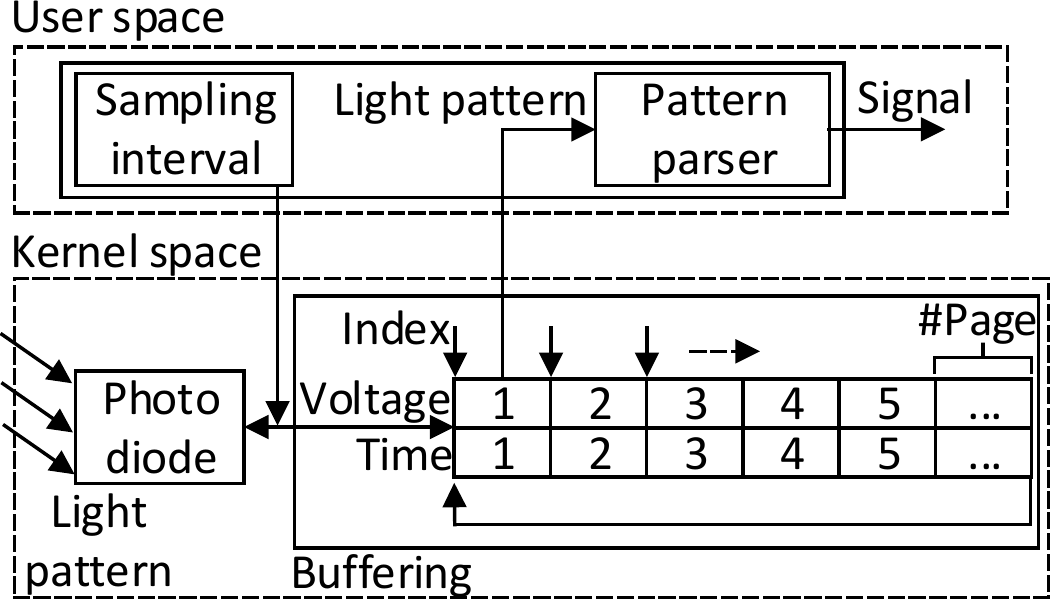}
			\caption{Scheme of light receiver \newline}
			\label{fig:light-receiver-scheme}
		\end{subfigure}
	\end{subfigure}
	\hfill
	\begin{subfigure}[b]{0.45\linewidth}
		\includegraphics[width=\linewidth]{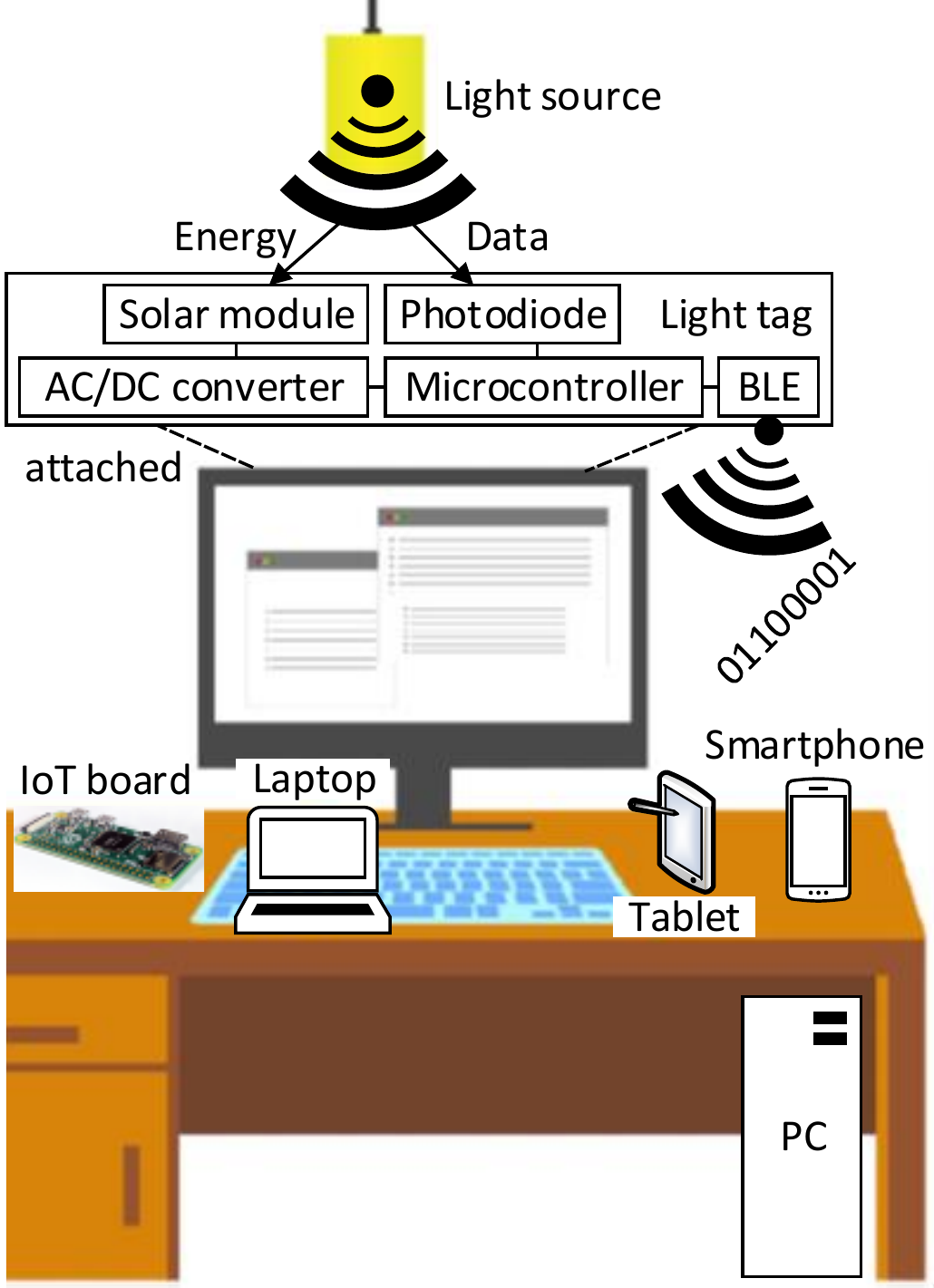}
		\caption{Light tag for ubiquitous visible light 
			signaling}
		\label{fig:vlc-tag}
	\end{subfigure}
	\caption{Implementation details of VLC sender and receiver 
	used by DevLoc. As future work we show the design of a light 
	tag for improved end-device support of VLC.}
\end{figure}

\subsection{Technical Details of DevLoc}
As system and development platform for our custom light bulb, 
we use the small, low-cost, single-board computer BeagleBone 
Black. We have implemented two Linux kernel modules to broadcast 
and receive light patterns. On the sender side, our custom light 
bulb broadcasts random light patterns via the light sender as 
shown in \prettyref{fig:light-sender-scheme}. The different 
light on and off periods of the signal pattern trigger two 
real-time kernel timers to switch the LED between on and off 
state. The custom light bulb consists of a power supply for the 
BeagleBone Black and during normal operation the battery is 
loaded and provides the power for the BeagleBone Black. The 
battery improves the service availability of DevLoc and 
maintains the lighting in case of a power blackout. In addition, 
the BeagleBone Black offers an API for visible light signaling 
and controls the LED transmitter and wireless modules such as 
Wi-Fi. On the receiving side, the light receiver in 
\prettyref{fig:light-receiver-scheme} samples the 
raw light signal via the photodiode. We have tested different 
sampling intervals, i.e., how often voltage values are sampled, 
and a sampling rate of 20\,µs works reliably to detect light 
patterns. This affects the signal buffering to store and access 
light signals from the kernel module. We save voltages and the 
relative time chunked into pages and control the maximum page 
size and number of pages based on the sampling rate to provide 
sufficient information for signal parsing and available system's 
memory.


We use MQTT for the communication among light bulbs. Via the 
subscription to the central backend, each master light bulb 
receives the configured light pattern which is then further 
published to the slave light bulb(s). Moreover, we use Python 
twisted as event-driven network programming framework to receive 
and send data between light bulbs and user clients. We establish 
a TLS network connection between the device grouping server and 
clients. The light bulb uses four different messages to query 
data for device grouping including raw light signal, detected 
light pattern, Wi-Fi, and Bluetooth scan data. The clients 
transmit the data in chunks of lines and the device grouping 
server buffers for each client the received raw data and merges 
them before the device association. The message format consists 
of a message type, a payload length, and the payload itself.

Regarding the VLC receiver, we aim to improve the end-device 
support for visible light signaling. Many user devices like 
personal computers do not have the required hardware such as a 
photodiode to sense light signals, while mobile user devices 
like 
smartphone and tablet provide a photodiode, e.g., for ambient 
light, but lack the ability to process light signals in 
real-time and require add-on hardware. For future work, we plan 
to shrink the light receiver to an appropriate (e.g., coin 
sized) volume for everyday usage towards ubiquitous visible 
light signaling. The foreseen light tag in 
\prettyref{fig:vlc-tag} acts as proximate communication hub 
which can be easily attached to different end-user devices 
enabling light transmissions. The light source acts as energy 
source and at the same time as transmission medium. Thereby, 
the light tag works passively meaning it awakes for operation 
via energy induced by the solar module. During operation the 
light tag receives and processes light transmitted data in 
real-time and broadcasts it via Bluetooth Low Energy (BLE) to 
nearby end-user devices.

\section{Evaluation of Device Associations via DevLoc}
We evaluate the propagation characteristics of VLC to qualify 
the feasibility of real-world deployments of DevLoc. Besides 
that, we emulate two varying environments with static and 
moving users to highlight the performance of DevLoc in different 
environments. We identify, for each case, the best working 
device association with regard to high detection accuracy and 
fast runtime. Therefore, we perform a thorough parameter 
estimation of our device grouping including the sampling periods 
for light patterns, device localization for comparison, and 
training classifiers. Moreover, we select the best performing 
distance and correlation metrics and determine the most suitable 
time-series features for light patterns.

\subsection{Propagation Characteristics of VLC}
With regard to the use of DevLoc in the real world, we have 
evaluated the maximum attainable range of light patterns for two 
different LEDs as VLC transmitter in a dark room without 
interference from the surrounding light \cite{Haus.2019b}. We 
used two LEDs: an omnidirectional LED with a weak light signal 
and a directional LED with a strong, beaming light signal. The 
directional LED reaches a maximum distance of 10\,m, while the 
omnidirectional LED can cover a distance of 3\,m. Furthermore, 
we identified via an experiment the FoV at the photodiode of the 
VLC receiver, the entire FoV ranges from 180° to 0°. The 
omnidirectional LED receives a range of 165°--50°, meaning opens 
at 165° and closes at 50°, and the directional LED achieves a 
FoV of 175°--5°. 
During our experiments, we have recognized that the impact of 
ambient light is decisive for VLC. Obviously, the directional 
LED is less sensitive to the ambient light compared to the 
omnidirectional LED. Nevertheless, with an active light source 
or direct sunlight acting as ambient light, the performance to 
detect signal patterns from the directional LED drops 
significantly. On the other hand, the omnidirectional LED only 
works reliably at a low ambient light intensity. For future 
work, we will adopt the algorithm in \cite{Guler.2018} which 
uses orthogonal codes to detect and isolate adjacent light 
sources. Thereby, we plan to enhance the robustness of DevLoc by 
supporting overlapping light patterns from different light bulbs.


\begin{figure}
	\centering
	\begin{subfigure}[b]{0.39\linewidth}
		\includegraphics[width=\linewidth]{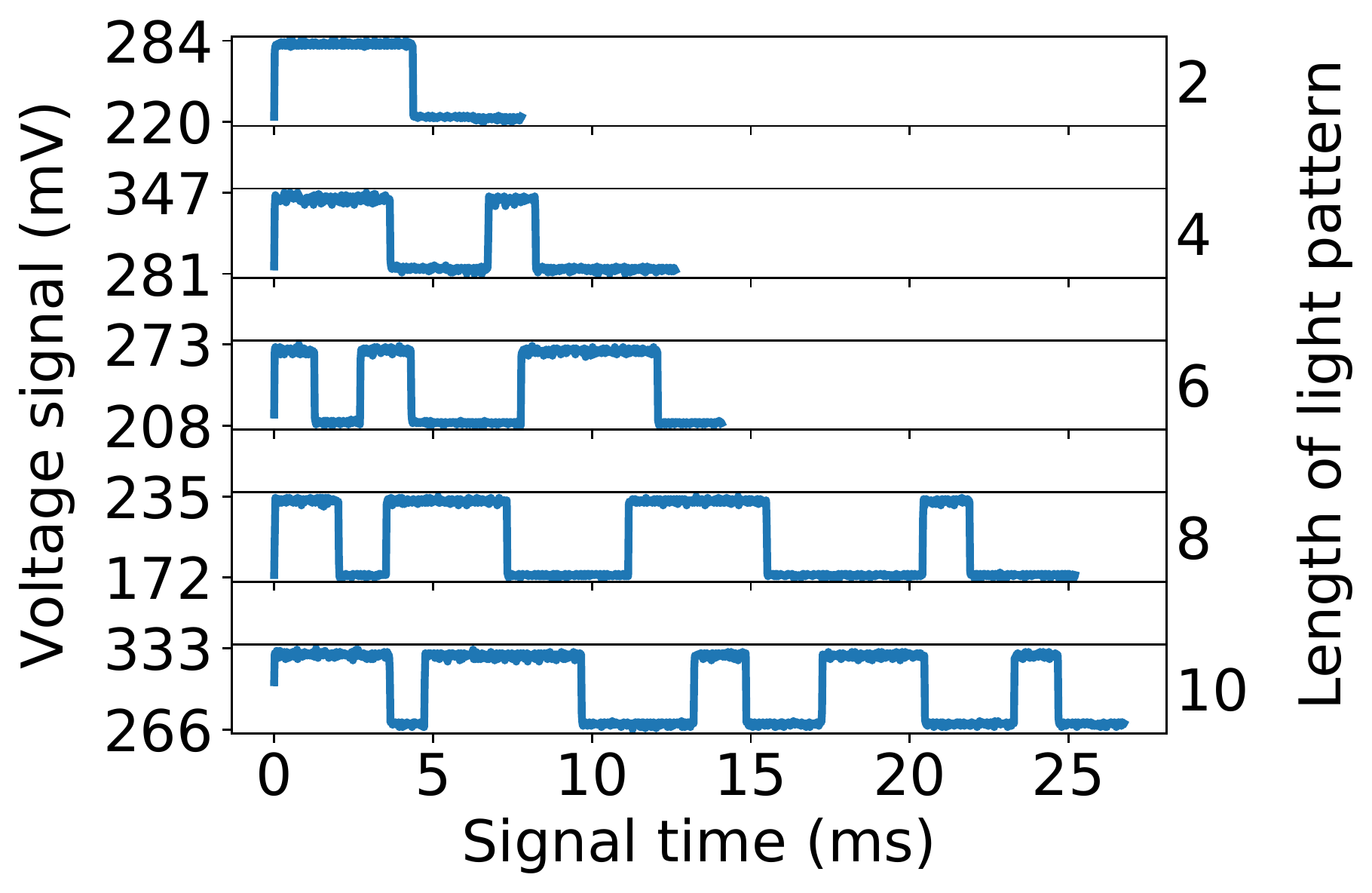}
		\caption{Random light patterns}
		\label{fig:random-light-pattern}
	\end{subfigure}
	\hfill
	\begin{subfigure}[b]{0.58\linewidth}
		\includegraphics[width=\linewidth]{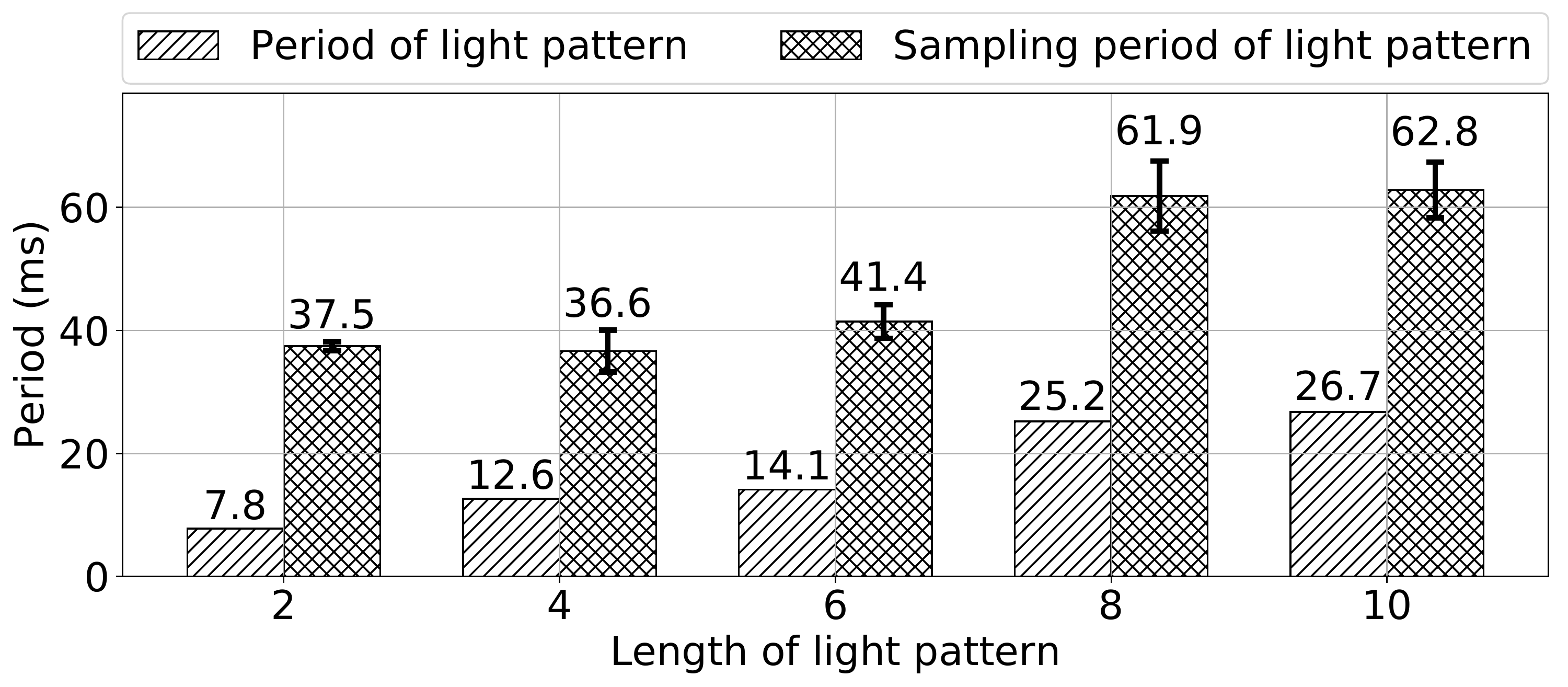}
		\caption{Average sampling period}
		\label{fig:sampling-period-light-patterns}
	\end{subfigure}
	\caption{Sampling periods to detect light patterns with a 
		varying length for device association}
\end{figure}

\subsection{Parameter Estimation:\\Sampling Periods of Light 
Patterns for Similarity Metrics}
Our device grouping takes advantage of random light patterns in 
different spatial areas to establish device groups for data 
sharing.
Which sampling periods are required to successfully detect light 
patterns? On the receiver side, we analyze the sampling time to 
be able to detect a valid light pattern within the time-series 
of voltage values. The light receiver 
is only able to recognize the light pattern when it starts 
repeating.
We use five different light patterns for our evaluation with a 
varying length of light on and off periods as shown in 
\prettyref{fig:random-light-pattern}. For each specific light 
pattern and over ten different test rounds, we choose a random 
start position within the light pattern and we extract a raw 
voltage signal via a monotonically increasing sampling time 
until all detected light 
patterns are valid. We classify a raw light signal as valid 
if all extracted light patterns have the same length $\in$ 
\{2, 4, 6, 8, 10\} and the duration of each light on and off 
phase is above 1\,ms known by the generation of light patterns. 
\prettyref{fig:sampling-period-light-patterns} shows 
the necessary sampling periods to successfully recognize a 
reoccurring light pattern compared to the duration of a single 
light pattern. The sampling time is on average 3.1 times longer 
than the raw signal pattern. In our evaluation we use the 
identified sampling ranges for each length of signal pattern 
to randomly choose a raw voltage signal which is large enough 
for a reasonable signal comparison.


\begin{figure}
	\centering
	\begin{subfigure}[b]{0.37\linewidth}
		\includegraphics[width=\linewidth]{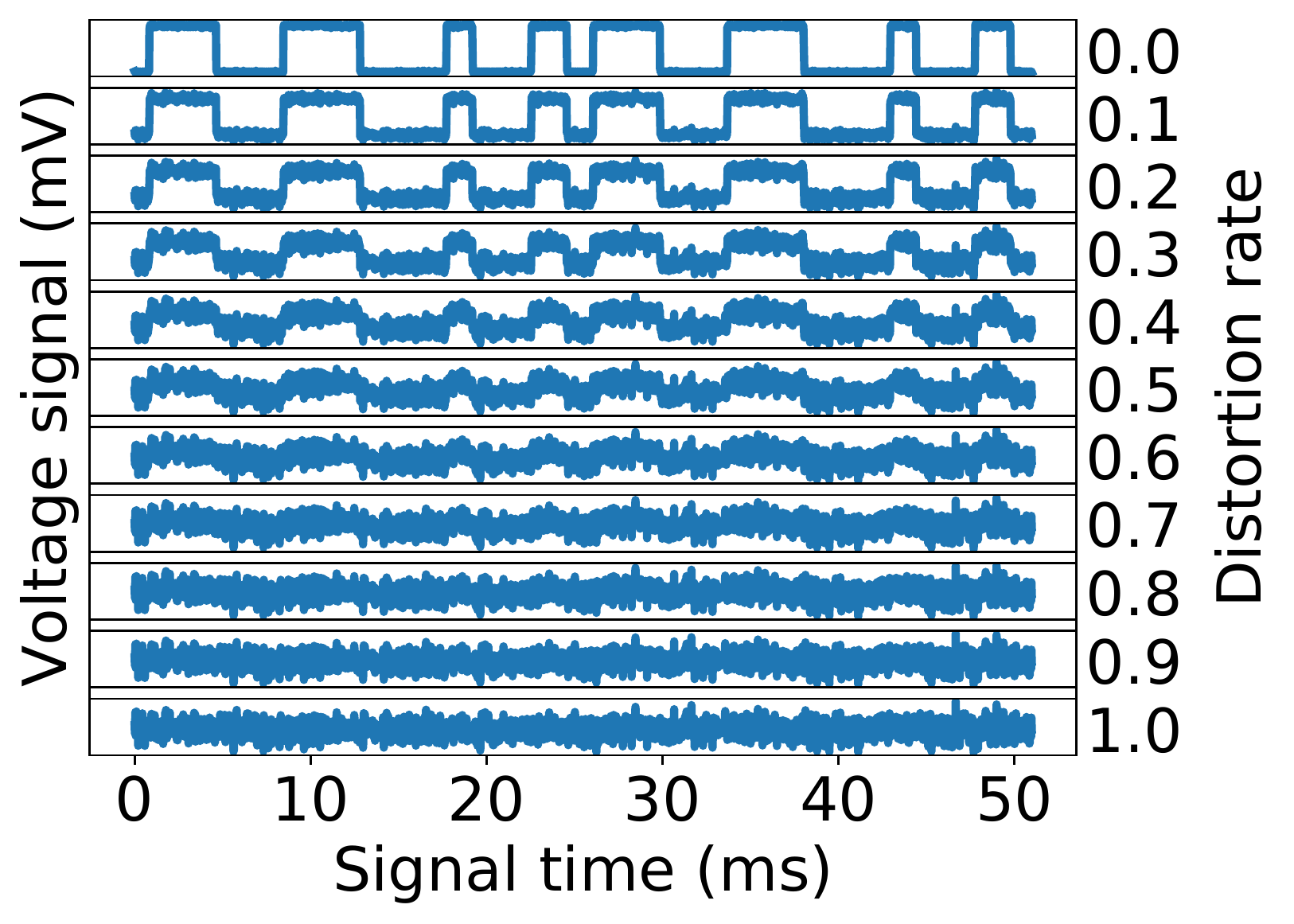}
		\caption{Distorted light signals}
		\label{fig:device-grouping-offline-distorted-signals}
	\end{subfigure}
	\hfill
	\begin{subfigure}[b]{0.61\linewidth}
		\includegraphics[width=\linewidth]{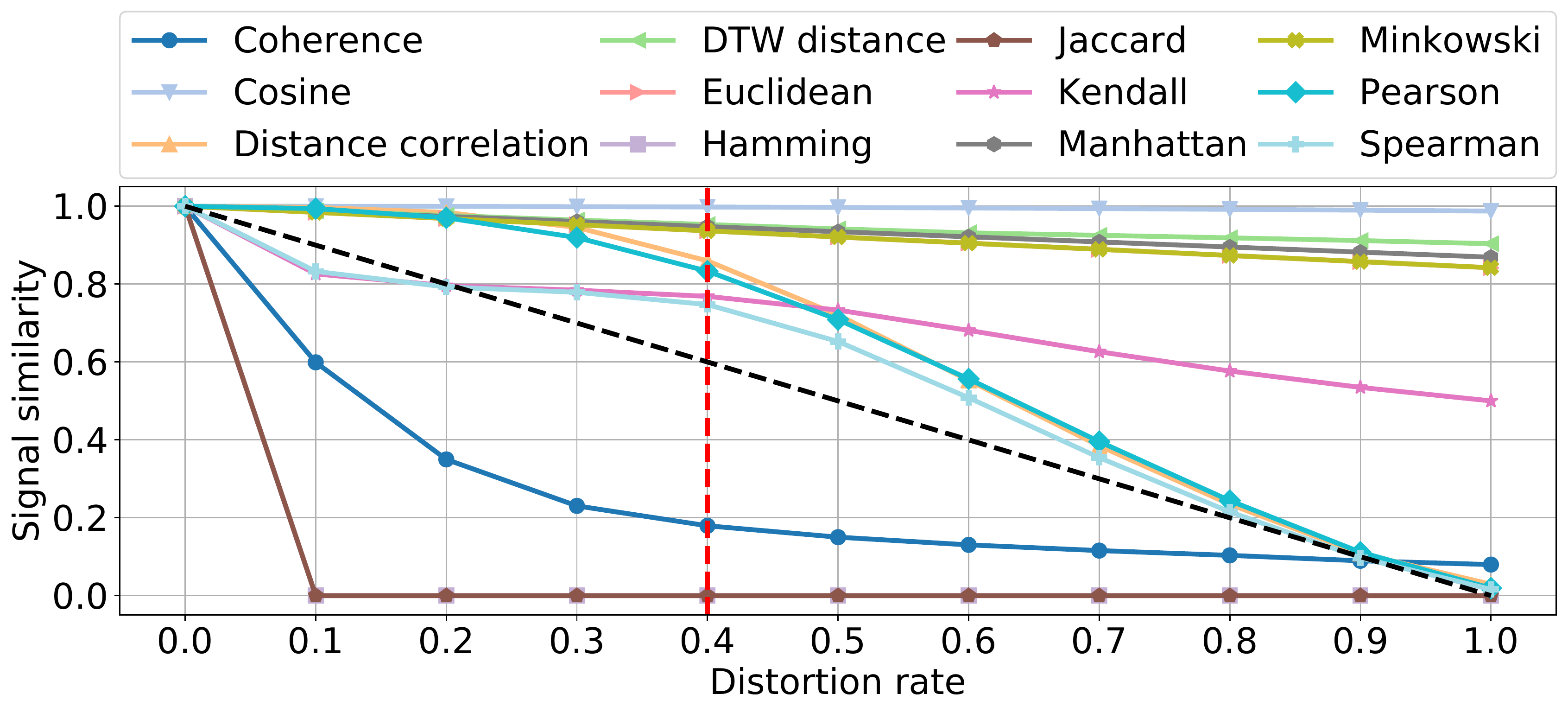}
		\caption{Signal similarity of distorted signals}
		\label{fig:device-grouping-offline-distortion-similarity}
	\end{subfigure}
	\caption{Analysis of signal similarity using distorted light 
		signals for several distance and correlation metrics}
\end{figure}

\begin{table}
	\caption{Best working similarity and equalize methods for 
	device association including similarity threshold and 
	runtime}
	\label{tab:device-grouping-offline-similarity}
	\centering
	\begin{tabular}{L{1.7cm}L{1cm}R{1.2cm}R{1.2cm}R{1.2cm}}
		\toprule
		Similarity measure &
		Equalize method &
		Average metric &		
		Similarity threshold &
		Runtime
		\\
		\midrule
		Pearson & DTW & 0.93 & 0.8 & 0.532\,s
		\\
		\midrule
		Spearman & DTW & 0.89 & 0.9 & 0.532\,s
		\\
		\midrule
		DTW distance & DTW & 0.89 & 0.7 & 0.506\,s
		\\
		\bottomrule
	\end{tabular}
\end{table}

\subsection{Parameter Estimation:\\Similarity Metrics for Light 
Patterns}
We identify the best working similarity metrics for light 
patterns in terms of highest accuracy for group detection. We 
analyze the behavior of the similarity metrics by comparing raw 
light signals with the same increasingly distorted light signal 
such as in 
\prettyref{fig:device-grouping-offline-distorted-signals}. 
Per similarity metric and signal distortion rate, the 
evaluation result in 
\prettyref{fig:device-grouping-offline-distortion-similarity} 
shows the median similarity over ten rounds and each light 
pattern. The desired property of the similarity metric is that 
the similarity decreases in case of increasing dissimilarity 
between two time-series signals. Hence, we identify the 
following reasonable similarity metrics at a signal distortion 
rate of 40\,\%, highlighted as dotted red line in 
\prettyref{fig:device-grouping-offline-distortion-similarity}: 
Spearman with a similarity threshold of 0.74, Pearson with a 
similarity threshold of 0.83, and distance correlation with a 
similarity threshold of 0.86.

Besides that, we simulate a testbed with two clients for device 
grouping where we perform ten evaluation rounds for each 
combination among all light patterns $\in$ \{2, 4, 6, 8, 10\} to 
find the best working similarity metric and equalize method to 
unify signal lengths being able to compare them. In each run, we 
apply the similarity metrics mentioned in 
\prettyref{fig:device-grouping-offline-distortion-similarity} 
and, as input, we randomly choose two light patterns and 
equalize their signal length using the methods $\in$ \{fill, 
cut, dynamic time warping (DTW)\}. 
\prettyref{tab:device-grouping-offline-similarity} 
presents the three best working combinations of similarity 
metric, equalize method, and signal threshold in terms of 
highest average metric including accuracy, precision, recall, 
and F1-score weighted with 0.8 and lowest runtime weighted with 
0.2. To calculate the result metrics, we assume that the two 
simulated clients are in the same region if the light signals 
have the same repeating pattern. We use these identified best 
working similarity measures to limit the runtime of our 
simulations for device grouping.


\begin{figure}
	\centering
	\includegraphics[width=\linewidth]{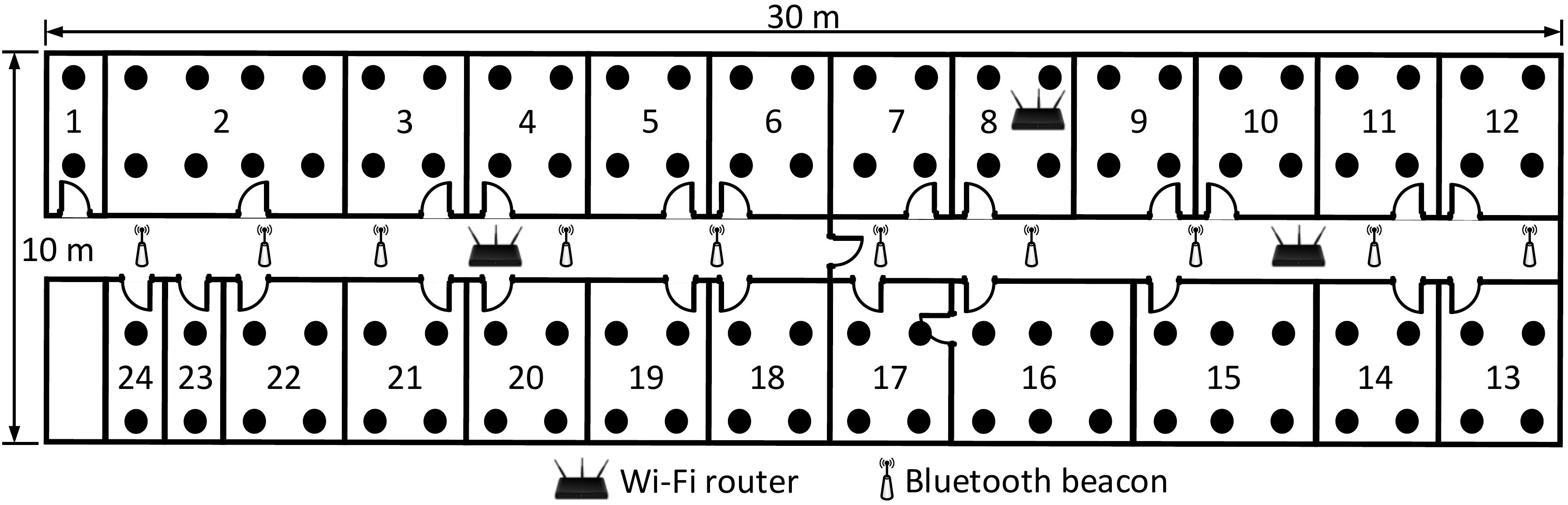}
	\caption{We model our university lab as simulation 
	environment for device associations. For comparison with 
	device localization, we take real traces of the Wi-Fi and 
	Bluetooth environment at different positions.}
	\label{fig:simulation-environment}
\end{figure}

\subsection{Parameter Estimation:\\Device Localization as 
Reference for Device Grouping}
We use well-known device localization as reference to compare 
the results of our device grouping based on light patterns. We
include a common indoor localization based on the similarity of 
Wi-Fi and Bluetooth signals \cite{Maghdid.2016}. Via an Android 
app we gather a list of Wi-Fi router and Bluetooth beacons 
containing MAC addresses and signal strengths (RSSI) for each 
measurement point ({\protect\tikz \protect\fill (0,0) circle 
(0.1);}) at our university lab as shown in 
\prettyref{fig:simulation-environment}.
We evaluate the sampling period, i.e., how many 
traces are required to achieve a 
reasonable localization accuracy. For a supervised machine 
learning approach with 10-fold cross validation, we have taken 
the following feature subset from the work in 
\cite{Sapiezynski.2016} to compare the list of Wi-Fi routers 
seen at two different measurement points, the same applies for 
the list of Bluetooth beacons:
\begin{itemize}
	\item Number of overlapping devices
	\item Size of the union of the two lists
	\item Jaccard distance between the size of the intersection 
	and the size of the union of the two lists	
	\item Number of non-overlapping devices
	\item Manhattan distance of RSSI of overlapping devices
	\item Euclidean distance of RSSI of overlapping devices
	\item Spearman correlation of RSSI of overlapping devices
	\item Pearson correlation of RSSI of overlapping devices
	\item Share top device based on strongest RSSI
	\item Share at least one top device based on RSSI range 
	$\pm$ 6\,dB
\end{itemize}

We merge the different measurements for each room and perform 
binary classification among all rooms with multiple sampling 
periods $\in$ [2, 5, 10, 15, 20, 25, 30]\,s applying 
content-based filtering, support vector machine (SVM), and 
random forest. The content-based filtering uses the shortest 
cosine distance among room measurements to identify the user's 
room. On average, the sampling period with 5\,s achieves the 
highest accuracy for Wi-Fi and Bluetooth localization.

\begin{figure}
	\centering
	\begin{subfigure}[b]{0.46\linewidth}
		\centering
		\includegraphics[width=0.8\linewidth]{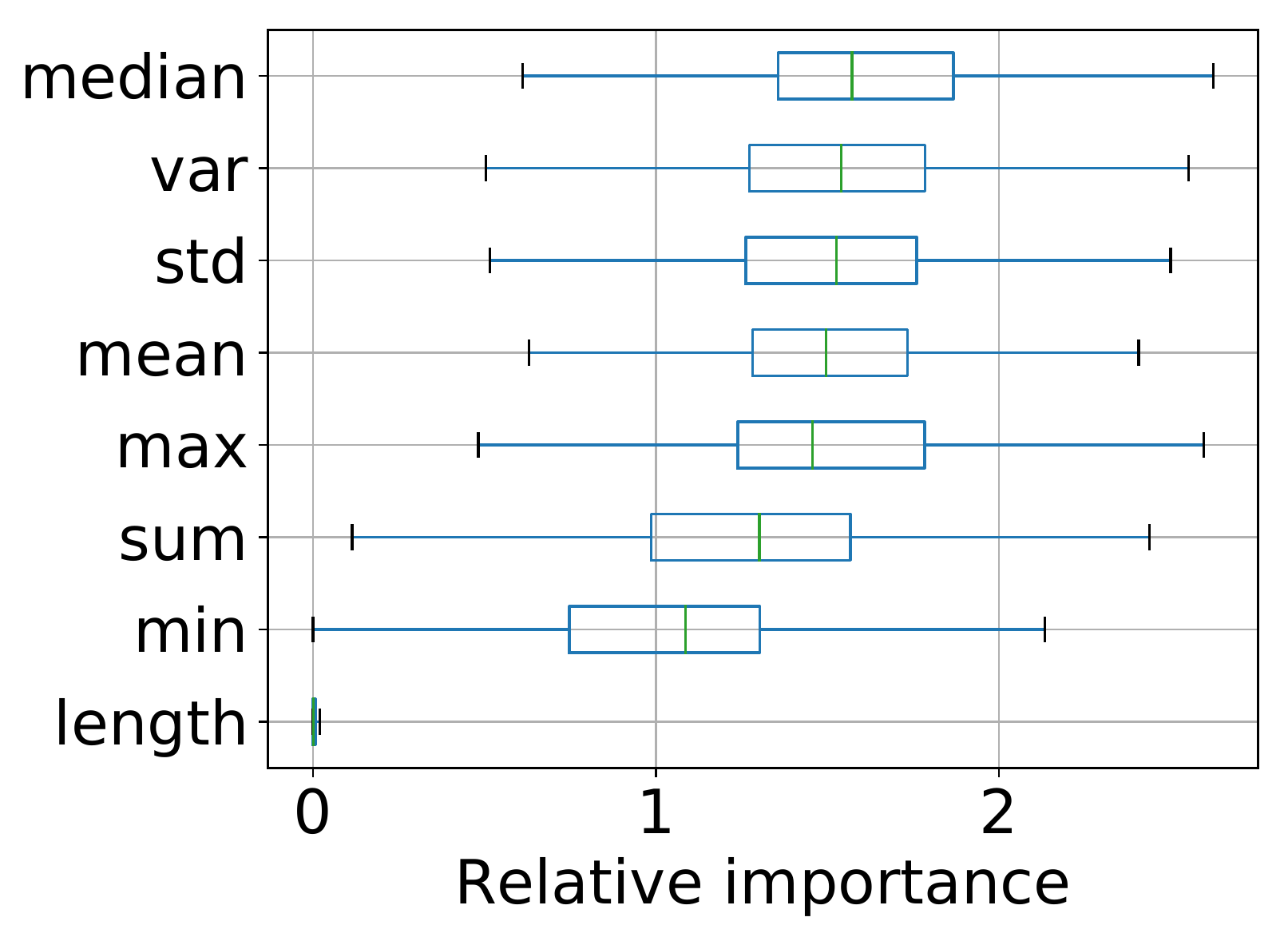}
		\caption{Single light patterns for dynamic 
		device-to-area simulation}
		\label{fig:feature-selection-basic-single}
	\end{subfigure}
	\hfill
	\begin{subfigure}[b]{0.52\linewidth}
		\centering
		\includegraphics[width=0.71\linewidth]{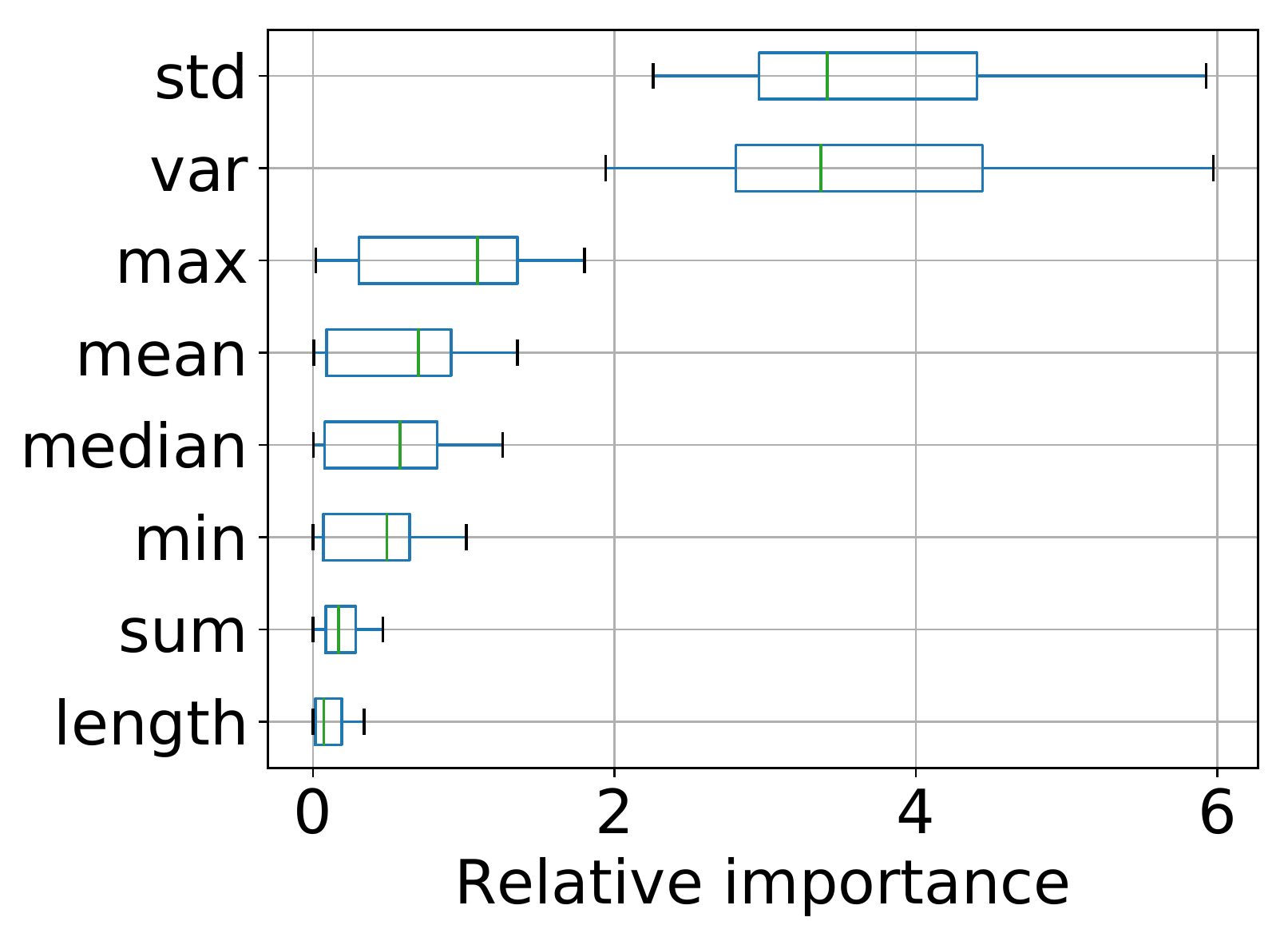}
		\caption{Combination of light patterns for static 
		device-to-device simulation}
		\label{fig:feature-selection-basic-combined}
	\end{subfigure}
	\caption{Selection of statistical features for light 
	patterns via relative importance of trained machine learning 
	models}
	\label{fig:feature-selection-basic}
\end{figure}

\vspace{5mm}

\subsection{Parameter Estimation:\\Feature Selection for 
ML-based Device Grouping}
Meaningful features are important for device grouping based on 
machine learning (ML) to achieve a good performance. Our feature 
selection identifies the features with highest entropy, i.e., 
information content, and lowest runtime. To find the most robust 
features in terms of distorted light patterns, we include light 
patterns with increasing white noise from 0\,\% to 100\,\%.

\textbf{Raw light patterns for different simulations} We use the 
combination of light patterns with different lengths for the 
static device-to-device simulation of our device grouping 
because it consists of only one room where we change the light 
pattern over time to keep the device groups up-to-date. Each 
master light bulb performs the proximity reasoning and is 
trained with a combination of light patterns with different 
lengths $\in$ \{2, 4, 6, 8, 10\}. In contrast, we use single 
light patterns for the dynamic device-to-area simulation of 
device grouping because it contains several rooms where the 
associated light pattern for each room remains the same over 
time. Each master light bulb is trained only for a specific, 
single light pattern with the same length.

\textbf{Feature types} We compute statistical features and 
time-series tailored features via tsfresh \cite{Christ.2018} for 
single and combination of light patterns. Tsfresh performs a 
time series feature extraction on basis of scalable hypothesis 
tests combining 63 time series characterization methods to 
identify the most meaningful features from a total of 794 time 
series features.



\begin{figure}
	\centering
	\begin{subfigure}[b]{0.46\linewidth}
		\centering
		\includegraphics[width=0.93\linewidth]{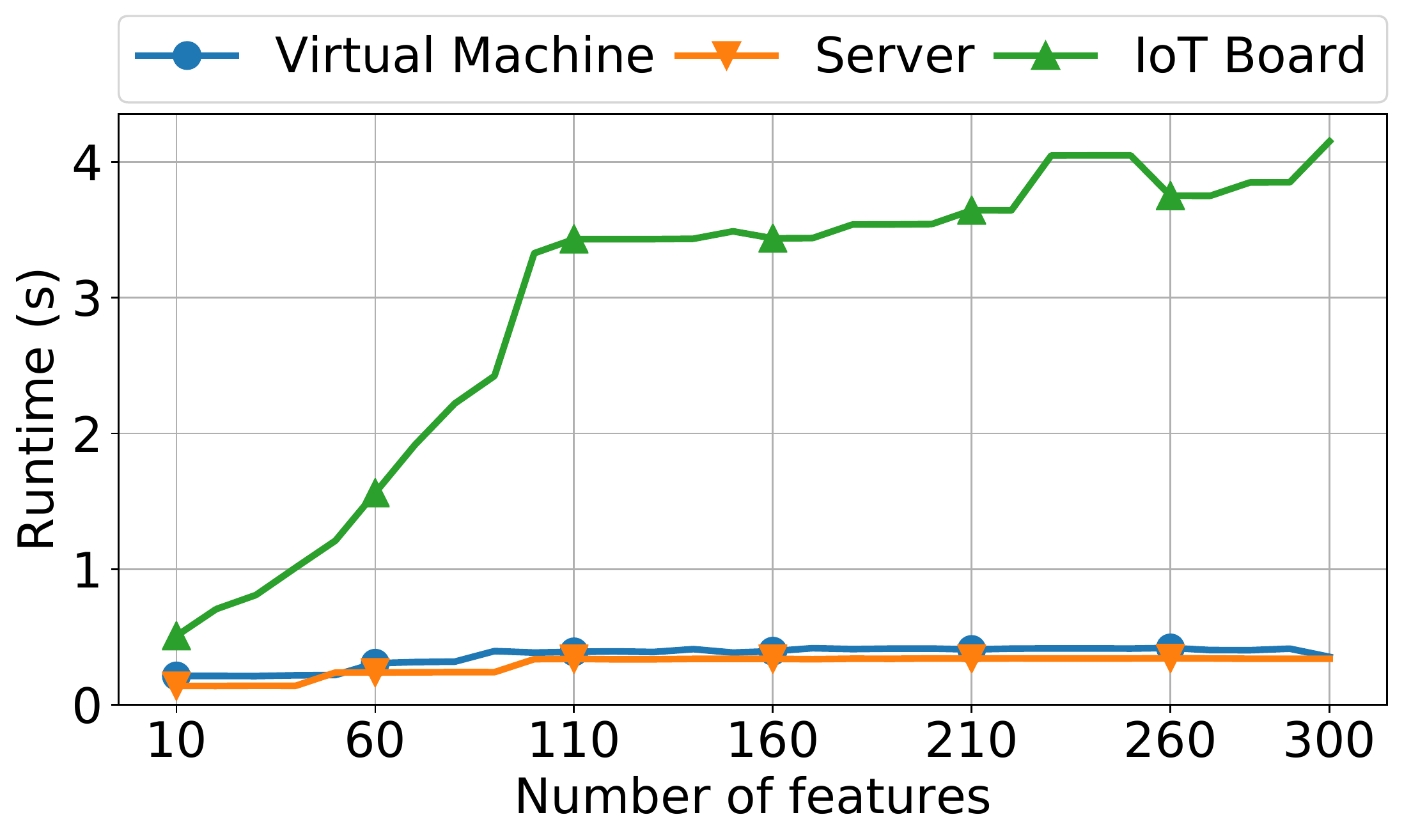}
		\caption{Single light patterns for dynamic 
			device-to-area simulation}
		\label{fig:feature-runtime-tsfresh-single}
	\end{subfigure}
	\hfill
	\begin{subfigure}[b]{0.52\linewidth}
		\centering
		\includegraphics[width=0.85\linewidth]{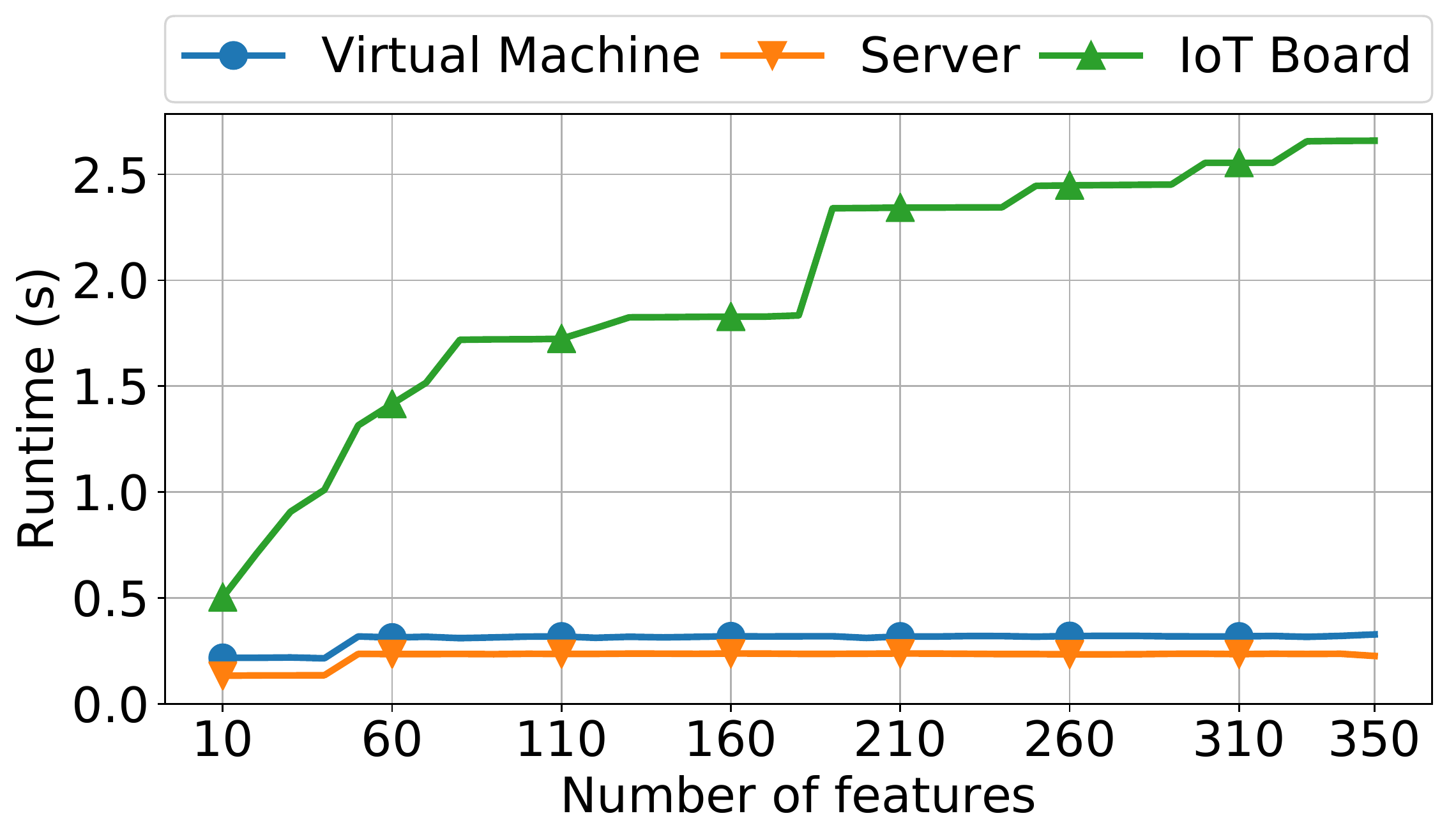}
		\caption{Combination of light patterns for static 
			device-to-device simulation}
		\label{fig:feature-runtime-tsfresh-combined}
	\end{subfigure}
	\caption{Runtime analysis of tsfresh features for light 
	patterns to highlight the performance difference among test 
	platforms}
	\label{fig:feature-runtime-tsfresh}
\end{figure}

\textbf{Best statistical features} For feature selection of 
statistical features we take advantage of three different 
machine learning models: extra trees, gradient boosting, and 
random forest to identify the most important features. 
\prettyref{fig:feature-selection-basic} shows the average 
relative importance of each statistical feature. In case of 
single light patterns, the relative importance is uniformly 
distributed over all features and only the feature length does 
not provide sufficient entropy. On the other hand, with the 
combination of light patterns, the variance and standard 
deviation outperforms all other features by 68\,\%. 

\begin{figure*}
	\centering
	\begin{subfigure}[b]{0.32\linewidth}
		\includegraphics[width=\linewidth]{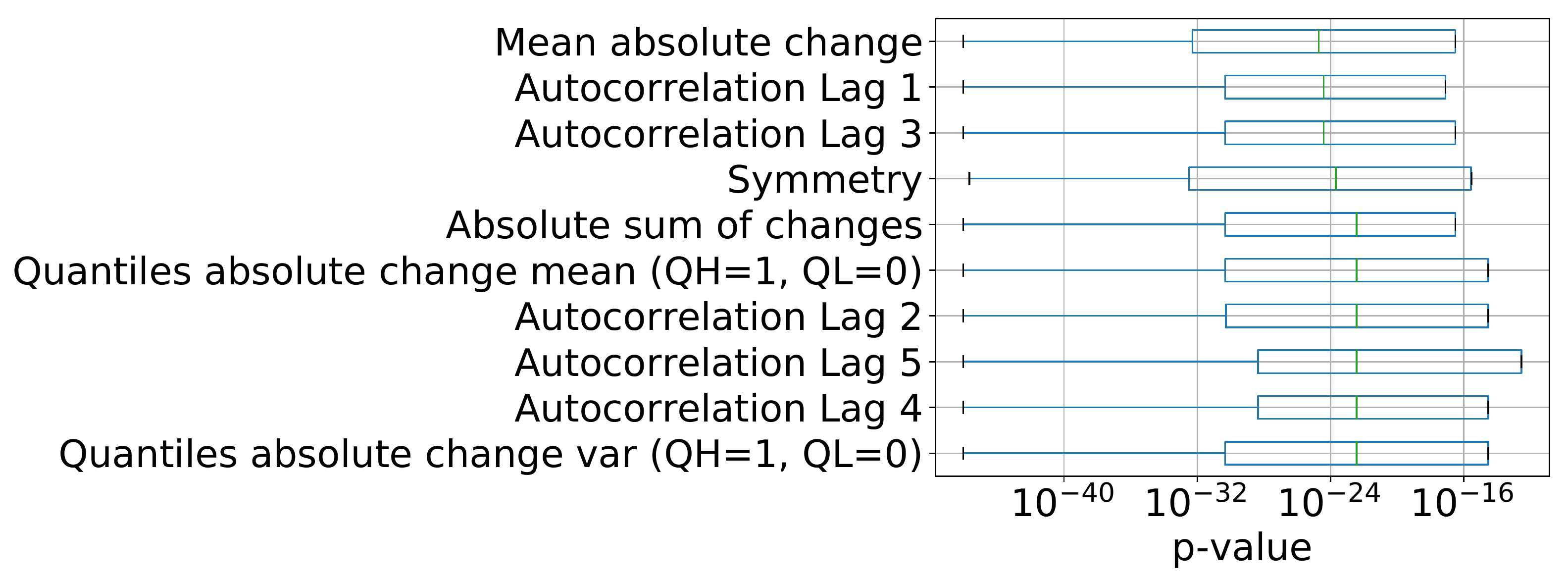}
		\caption{Best ten features of single light patterns for 
		dynamic device-to-area simulation}
		\label{fig:feature-selection-tsfresh-single}
	\end{subfigure}
	\hfill
	\begin{subfigure}[b]{0.32\linewidth}
		\includegraphics[width=\linewidth]{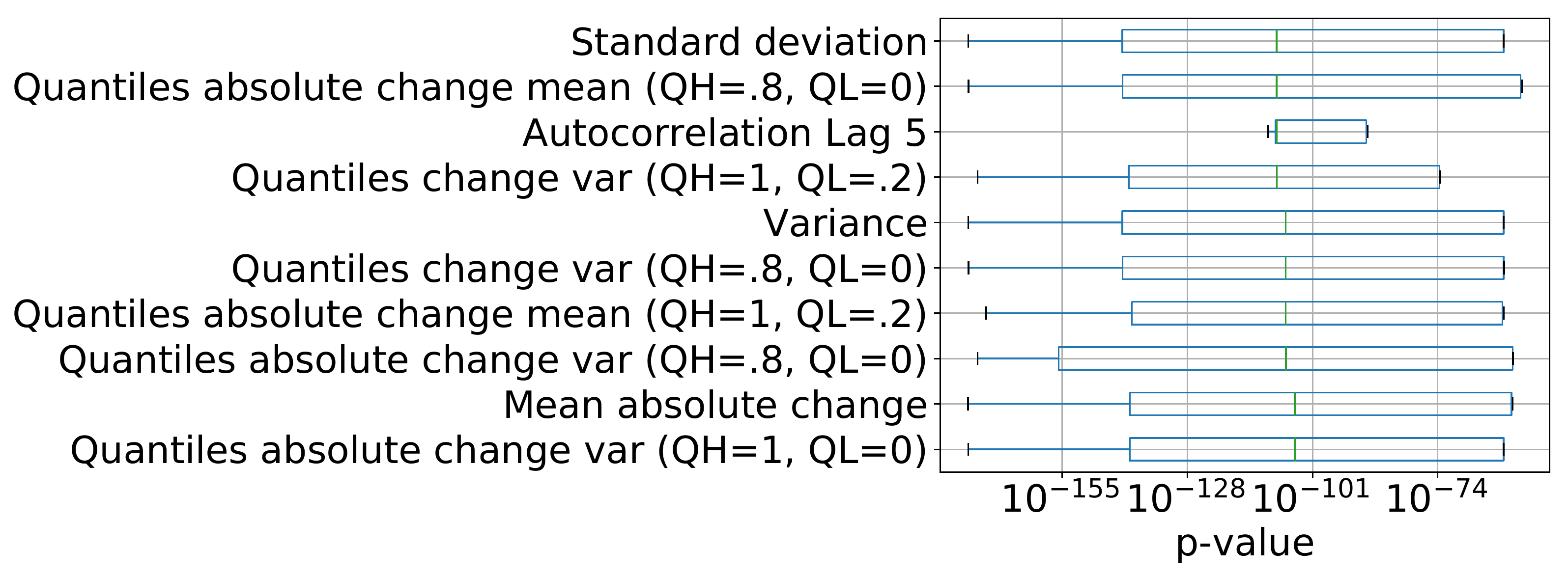}
		\caption{Best ten features for combination of light 
		patterns for static device-to-device simulation}
		\label{fig:feature-selection-tsfresh-combination}
	\end{subfigure}
	\hfill
	\begin{subfigure}[b]{0.3\linewidth}
		\centering
		\includegraphics[width=0.47\linewidth]{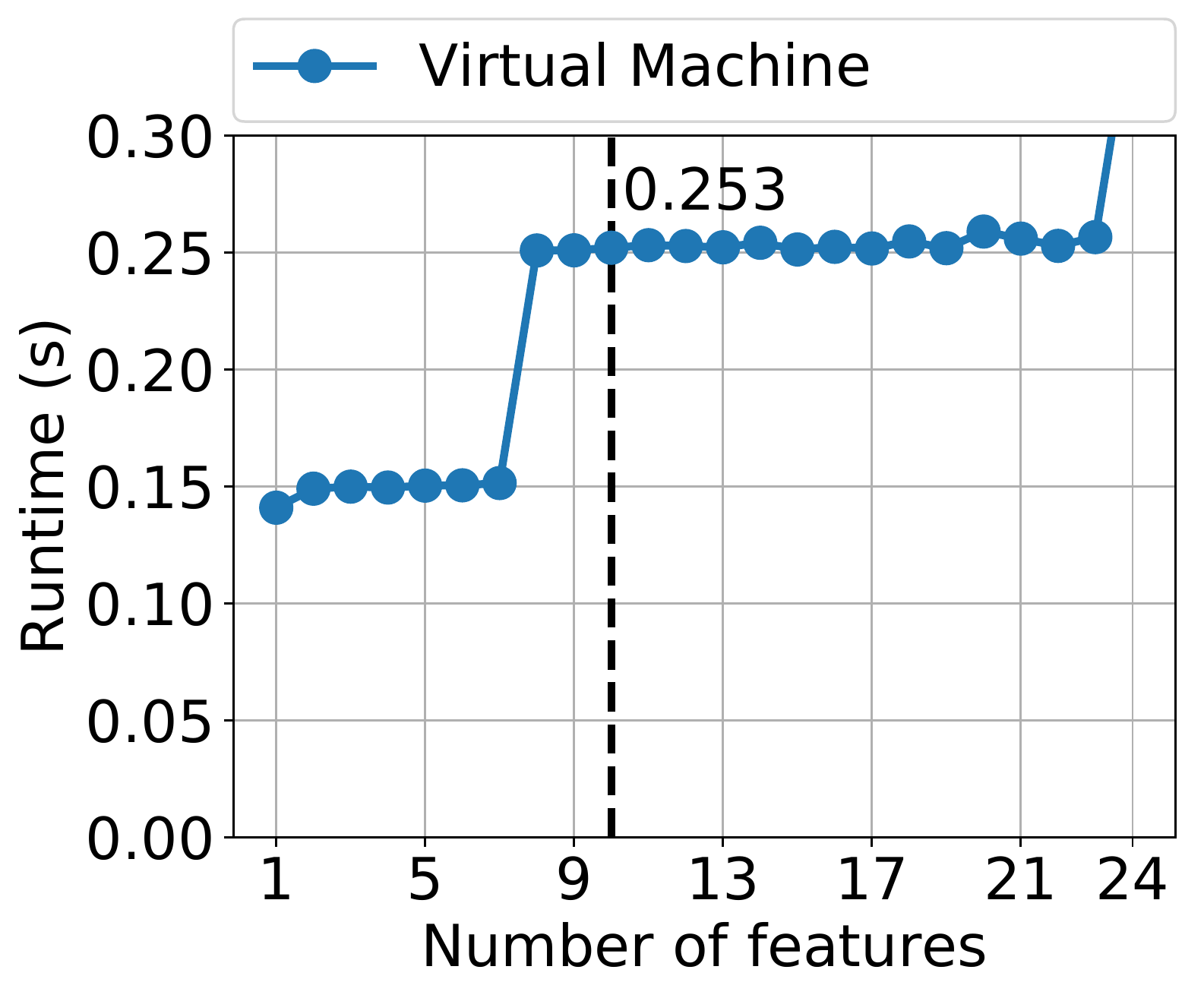}
		\caption{Detailed runtime analysis of best time-series 
		features at our simulation platform}
		\label{fig:feature-runtime-tsfresh-selected}
	\end{subfigure}
	\caption{Via tsfresh we select the ten most meaningful 
	features of each light pattern type for device grouping. 
	Moreover, we highlight the runtime to calculate these 
	features at the virtual machine used as simulation platform 
	for device grouping.}
	\label{fig:feature-selection-tsfresh}
\end{figure*}


\begin{table}
	\centering
	\caption{Testbed specifications for runtime analysis of 
	feature selection used by device grouping}
	\label{tab:hardware-testbeds}
	\begin{tabular}{L{2.7cm}ll}
		\toprule
		System &
		CPU &
		RAM
		\\
		\midrule
		Server & 40x Intel Xeon E5-2630 2.2\,GHz & 768\,GB 
		\\
		\midrule
		Virtual machine (VM) & 4x Intel Xeon E5-2630 2.2\,GHz 
		& 32\,GB
		\\
		\midrule
		Next unit of computing PC (NUC) & 4x Intel Core i5-6260U 
		1.8\,GHz & 16\,GB
		\\
		\midrule
		IoT board & 1x ARM AM3358 1\,GHz & 0.5\,GB
		\\
		\bottomrule
	\end{tabular}
\end{table}

\textbf{Runtime analysis of time-series features to select 
simulation platform} We apply tsfresh for time-series feature 
selection to identify the most meaningful features of light 
patterns. Thereby, we highlight the runtime of feature 
calculation to evaluate the practicality of our proposed system 
where we aim to run the device associations directly at the 
light bulb which embeds an IoT board with limited hardware 
capabilities. \prettyref{fig:feature-runtime-tsfresh} presents 
the feature runtime for single and combination of light patterns 
on three different test platforms described in 
\prettyref{tab:hardware-testbeds}. We sort the features 
according to their decreasing information content based on the 
hypothesis testing from tsfresh. For runtime comparison we 
calculate the relative runtime = 
\nicefrac{\text{runtime}}{\text{number of features}} among the 
test platforms. In case of the single light patterns, tsfresh 
computes 300 features where the server achieves the fastest 
performance with a relative runtime of 2.63\,ms per feature, the 
virtual machine is about 31\,\% slower with 3.46\,ms per 
feature, and the IoT board is 6.2 times slower with 21.28\,ms 
per feature. Using the combination of light patterns we achieve 
a similar runtime where tsfresh calculates 350 different 
time-series features. The server reaches a relative runtime of 
2.11\,ms per feature, the virtual machine with 3.17\,ms, and the 
IoT board is by far the slowest platform with 14.99\,ms per 
feature. Based on our findings from the feature runtime, we 
choose the virtual machine (VM) as simulation platform to 
evaluate DevLoc because the IoT board as desired platform is too 
slow when we repeat the device grouping at a higher rate. Note 
that this is just for the purpose of evaluation, the IoT board 
is still considered in other parts of the evaluation.

\textbf{Best time-series features} To improve the detection 
accuracy and limit the runtime of our device grouping, we select 
the ten most meaningful features of light patterns which are 
shown in Figs. \ref{fig:feature-selection-tsfresh-single} and 
\ref{fig:feature-selection-tsfresh-combination}. We use the 
p-value or probability value from tsfresh to select the most 
important features, the probability of finding the observed 
results when the null hypothesis is true. The lower the p-value 
the more significant is the feature. To ensure a reasonable 
performance for device grouping, 
\prettyref{fig:feature-runtime-tsfresh-selected} 
presents a detailed runtime analysis of features of light 
patterns computed at the virtual machine used as simulation 
platform for device grouping. It takes 253\,ms to compute the 
ten most meaningful features for device associations which is 
fast enough to ensure the validity of our simulation results 
considering that we repeat device grouping every few seconds 
among multiple users.


\subsection{Parameter Estimation: Sampling Period of Light 
Patterns to Train ML-based Device Grouping}
\begin{figure}
	\centering
	\begin{subfigure}[b]{0.322\linewidth}
		\includegraphics[width=\linewidth]{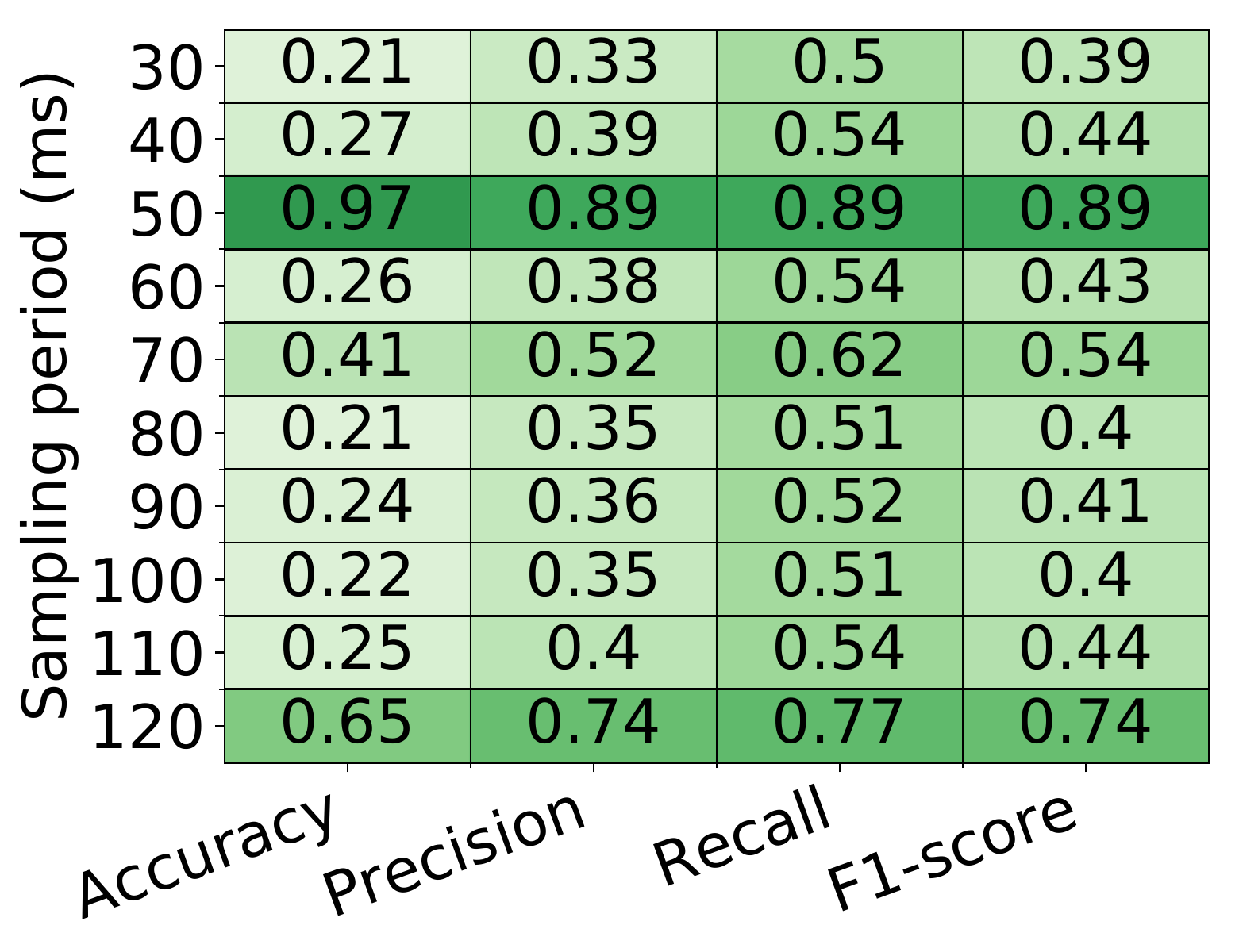}
		\caption{Extra trees}
	\end{subfigure}
	\hfill
	\begin{subfigure}[b]{0.322\linewidth}
		\includegraphics[width=\linewidth]{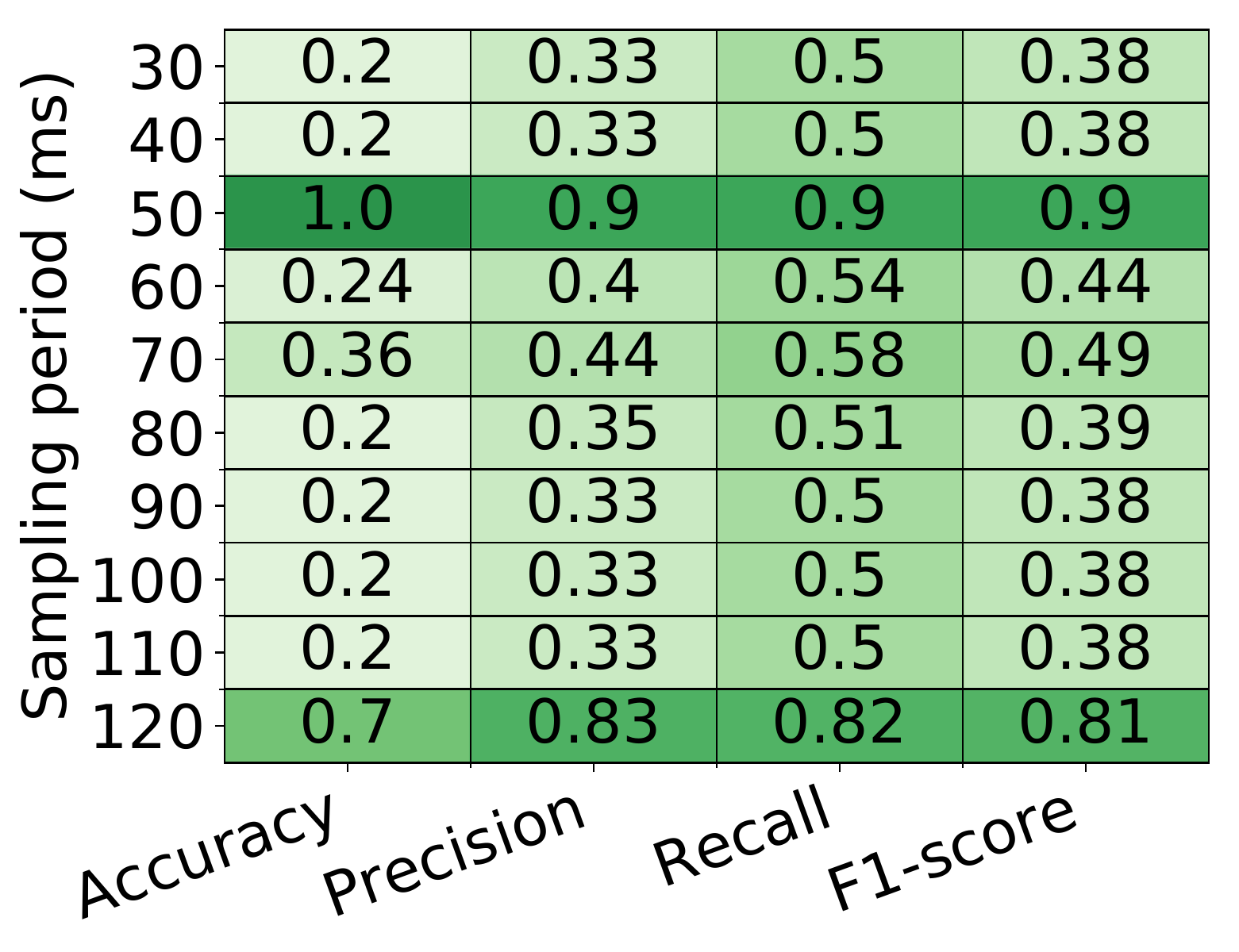}
		\caption{Gradient boosting}
	\end{subfigure}	
	\hfill
	\begin{subfigure}[b]{0.322\linewidth}
		\includegraphics[width=\linewidth]{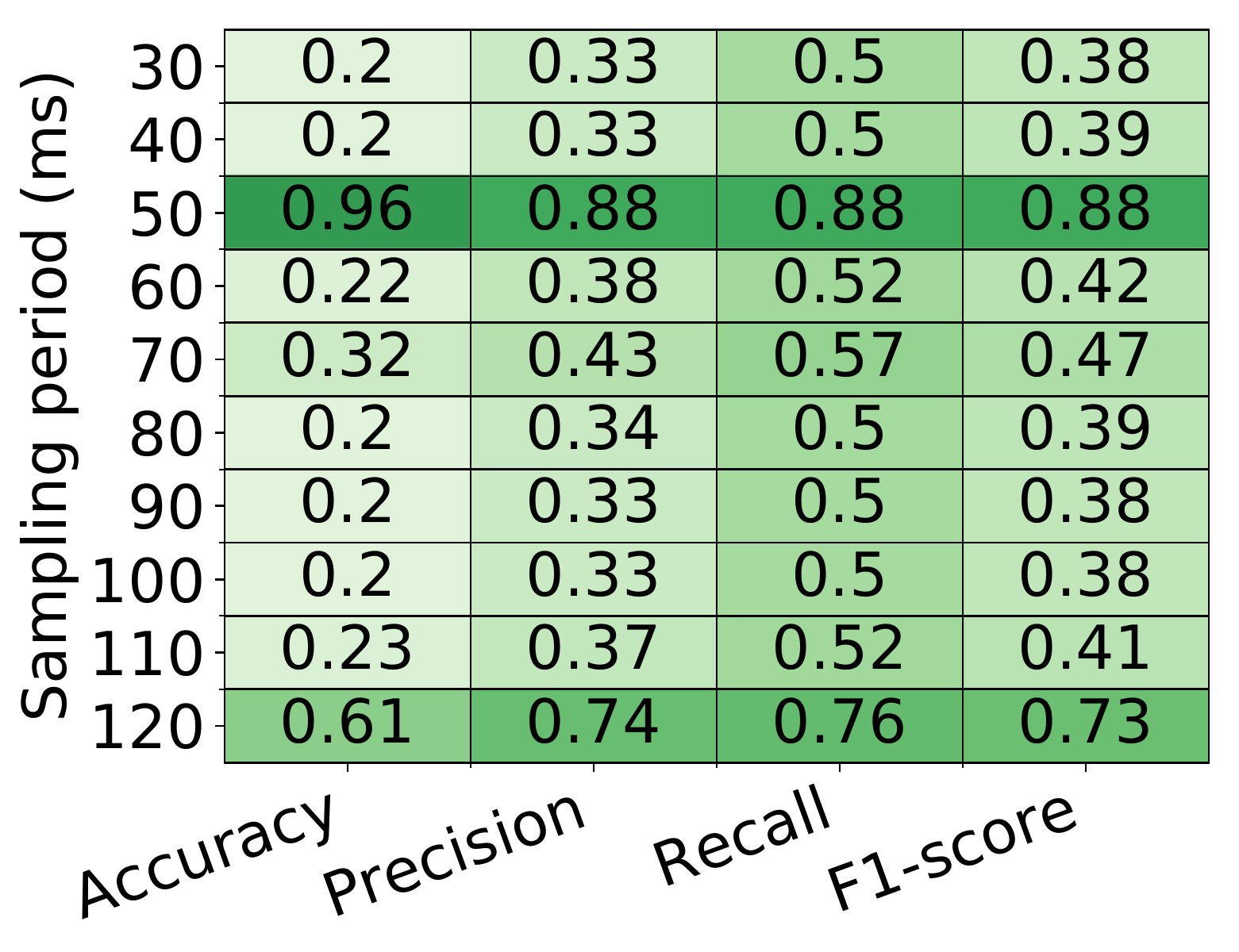}
		\caption{Random forest}
	\end{subfigure}
	\caption{Parameter estimation of sampling period to train 
	ML-based device association}
	\label{fig:ml-offline-params}
\end{figure}

We perform an offline evaluation of our device grouping to 
identify the best working sampling period for light patterns to 
train different classifiers. In our experiment, the device 
association ranges between two and ten grouping clients and we 
apply 10-fold cross validation for the classifiers: extra trees, 
gradient boosting, and random forest. These are trained via 
sampling periods $\in$ [30, 120]\,ms for different light 
patterns with signal length $\in$ \{2, 4, 6, 8, 10\} and 
selected tsfresh features from 
\prettyref{fig:feature-selection-tsfresh}. We take the average 
results of our ML-based device grouping including accuracy, 
precision, recall, and F1-score. 
\prettyref{fig:ml-offline-params} shows that the sampling 
period of 50\,ms achieves the best result with 0.91 over all 
classifiers compared to a sampling period of 120\,ms with 0.74. 
The classifiers: extra trees, gradient boosting, and random 
forest achieve similar results.




\subsection{Simulation Parameters for Device Grouping}
To evaluate DevLoc, we run two different simulations with static 
and dynamic users based on a dedicated simulator. 
\prettyref{tab:parameters-device-grouping} shows the summarized 
best working parameters for device grouping (italicized). With 
persisted environmental data from our university lab as shown in 
\prettyref{fig:simulation-environment}, we perform a 
trace-driven simulation where each grouping client uses three 
different real traces: Wi-Fi and Bluetooth scans, and random 
light patterns with varying length. To achieve a realistic 
simulation, we emulate the network latency between the grouping 
server and the clients. Each client waits a random time within a 
predefined time range before sending the requested environment 
data to the grouping server. Thereby, we select a random start 
time within the sensing range for light patterns, Wi-Fi and 
Bluetooth scans. In addition, we randomly choose a sampling 
period within the identified best working sampling ranges.

\begin{table}
	\centering
	\caption{Summarized settings from parameter estimation 
	(highlighted in italic) and simulation parameters 
	(highlighted in bold) for device grouping.
	}
	\label{tab:parameters-device-grouping}
	\begin{tabular}{lL{3.3cm}L{3.2cm}}
		\toprule
		Simulation & \multicolumn{2}{c}{Parameters}
		\\
		\midrule
		\multirow{2}{*}{\shortstack[l]{Static \&\\ Dynamic}} & 
		\textit{Sampling period to train similarity classifiers} 
		& 50\,ms
		\\
		\cmidrule{2-3}
		& \textbf{Similarity classifiers} & Random forest, extra 
		trees, gradient boosting
		\\
		\cmidrule{2-3}
		& \textit{Sampling period localization} & 5\,s
		\\
		\cmidrule{2-3}
		& \textbf{Localization classifiers} & Content-based 
		filtering, random forest, SVM
		\\
		\cmidrule{2-3}
		& \textit{Similarity equalize method} & DTW
		\\
		\cmidrule{2-3}
		& \textit{Similarity threshold} & 0.7
		\\
		\cmidrule{2-3}
		& \textbf{Similarity metrics} & Pearson, Spearman
		\\
		\midrule
		Dynamic & \textbf{Rooms} & [1, 2, 3, 4, 5, 6, 7, 8, 9, 
		10]
		\\
		\cmidrule{2-3}
		& \textbf{Users} & [3, 5, 10]
		\\
		\cmidrule{2-3}
		& \textbf{Grouping frequency} & [10, 20, 30]\,s
		\\
		\midrule
		Static & \textbf{Users} & [2, 3, 4, 5, 6, 7, 8, 9, 10]
		\\
		\cmidrule{2-3}
		& \textbf{Light patterns} & [2, 4, 6, 8, 10]
		\\
		\bottomrule
	\end{tabular}
\end{table}




\begin{table*}
	\centering
	\caption{We identify best working classifiers and features 
	for device grouping via simulations with static and moving 
	users}
	\label{tab:results-grouping}
	\begin{tabular}{$l^l^l^c^c^c^c^c^c}
		\toprule
		Simulation &
		Grouping technique &
		Feature type &
		Result &
		Runtime &
		Accuracy &
		Precision &
		Recall &
		F1-score
		\\
		\midrule
		Dynamic & Extra trees & Statistical & .83 & 0.26\,s (1) 
		& .75 & 1 & .75 & .83
		\\
		\cmidrule{2-9}
		& Content-based filtering & Wi-Fi & .84 & 0.61\,s (7) & 
		1 & .78 & .78 & .78
		\\
		\cmidrule{2-9}
		& Content-based filtering & Bluetooth & .84 & 0.59\,s 
		(6) & .95 & .81 & .81 & .81
		\\
		\cmidrule{2-9}
		& Gradient boosting & Selected statistical & .93 & 
		0.53\,s (5) & .9 & 1 & .9 & .93
		\\
		\cmidrule{2-9}
		& Gradient boosting & Selected tsfresh & .93 & 1.58\,s 
		(8) & .9 & 1 & .9 & .93
		\\
		\cmidrule{2-9}
		\rowstyle{\bfseries} & Pearson & Light signal & .95 & 
		0.28\,s (2) & .95 & .95 & .95 & .95
		\\
		\cmidrule{2-9}
		& Pearson & Light pattern & .95 & 0.47\,s (4) & 1 & .93 
		& .93 & .93
		\\
		\cmidrule{2-9}
		& Pearson & Duration of light pattern & .95 & 0.46\,s 
		(3) & 1 & .93 & .93 & .93
		\\
		\midrule
		Static & SVM & Wi-Fi & .31 & 2.61\,s (6) & .32 & .2 & 
		.44 & .26
		\\
		\cmidrule{2-9}
		& Random forest & Bluetooth & .43 & 2.64\,s (7) & .34 & 
		.48 & .52 & .38
		\\
		\cmidrule{2-9}
		& Pearson & Light signal & .64 & 2.06\,s (5) & .25 & .77 
		& .79 & .76
		\\
		\cmidrule{2-9}
		& Spearman & Duration of light pattern & .81 & 0.42\,s 
		(1) & 1 & .75 & .75 & .75
		\\
		\cmidrule{2-9}
		& Spearman & Light pattern & .81 & 0.49\,s (3) & 1 & .75 
		& .75 & .75
		\\
		\cmidrule{2-9}
		\rowstyle{\bfseries} & Gradient boosting & Statistical & 
		.81 & 0.45\,s (2) & 1 & .75 & .75 & .75
		\\
		\cmidrule{2-9}
		& Gradient boosting & Selected statistical & .83 & 
		0.75\,s (4) & .89 & .81 & .81 & .8
		\\
		\cmidrule{2-9}
		& Gradient boosting & Selected tsfresh & .96 & 
		4.02\,s (8) & 1 & .94 & .94 & .94
		\\
		\bottomrule
	\end{tabular}
\end{table*}

\begin{figure*}
	\centering
	\begin{subfigure}{0.3\linewidth}
		\centering
		\includegraphics[width=0.7\linewidth]{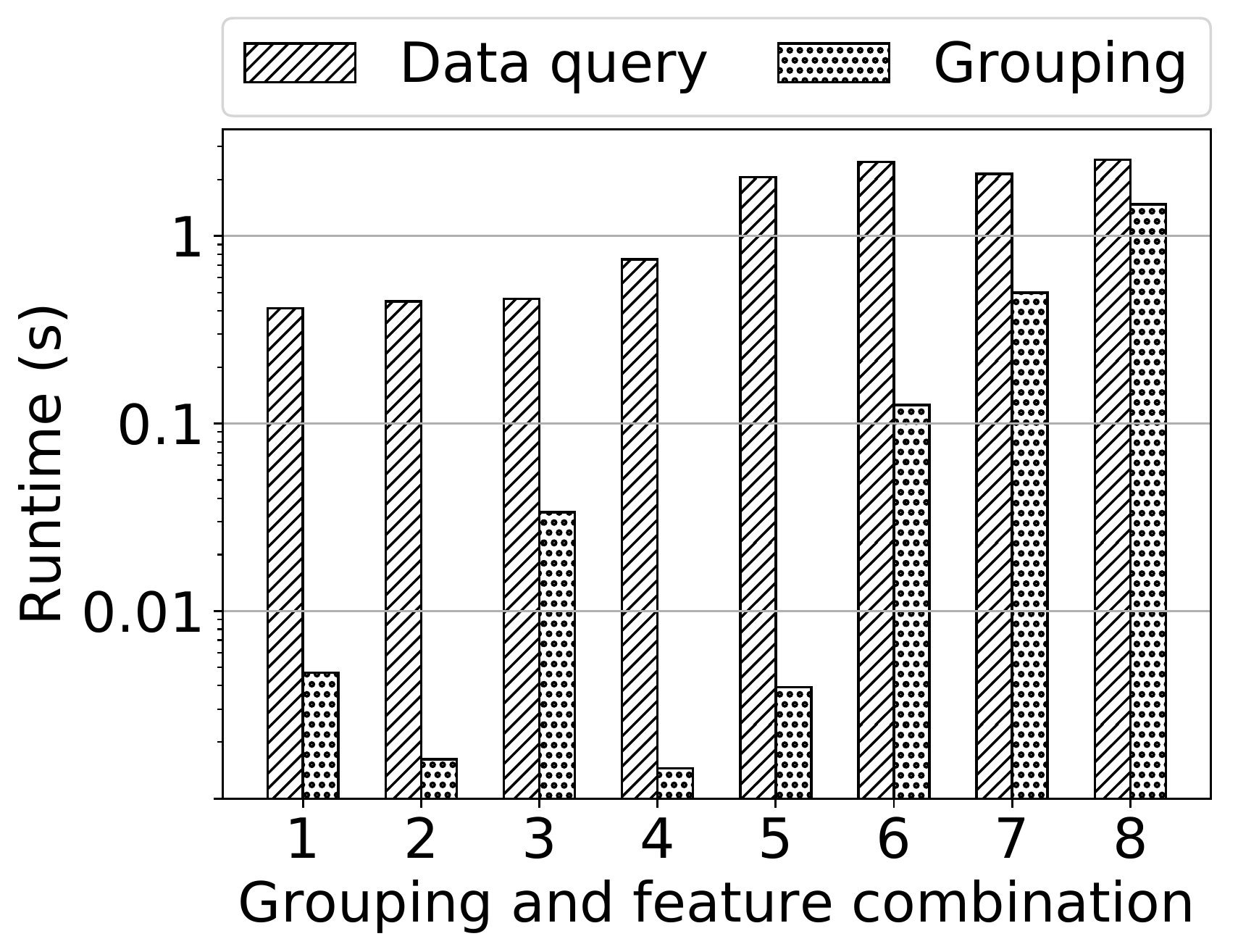}
		\caption{Static device-to-device grouping}
		\label{fig:runtime-static-coupling}
	\end{subfigure}
	\qquad
	\begin{subfigure}{0.3\linewidth}
		\centering
		\includegraphics[width=0.7\linewidth]{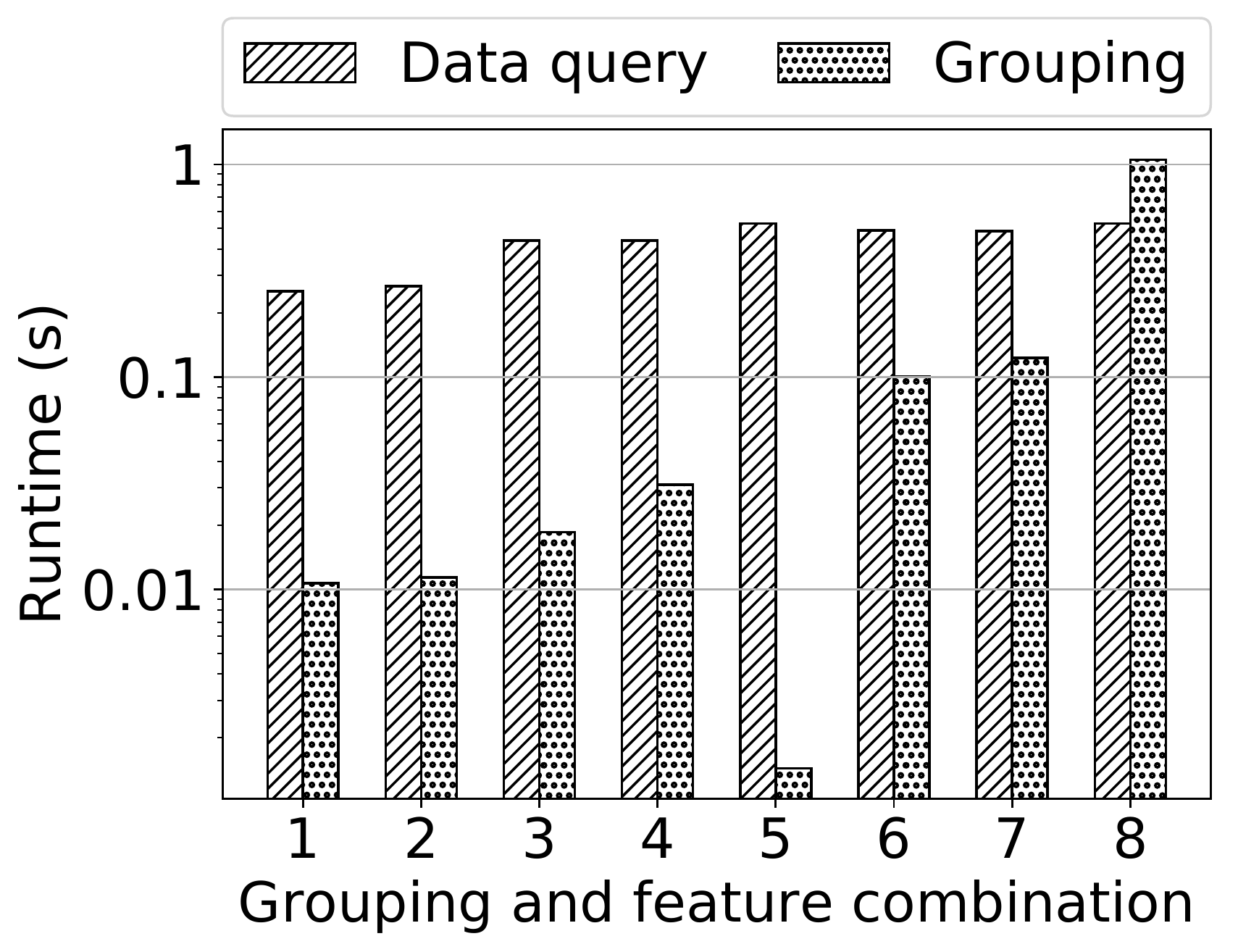}
		\caption{Dynamic device-to-area grouping}
		\label{fig:runtime-dynamic-coupling}
	\end{subfigure}
	\caption{Runtime analysis of different device groupings and 
	signal features. To identify the grouping technique and 
	feature type, the number of grouping and feature combination 
	matches with the runtime order in 
	\prettyref{tab:results-grouping} mentioned in brackets.}
\end{figure*}

\subsection{Static Device-to-Device Simulation of Device 
Grouping}

\textbf{Simulation settings} No user in the static simulation 
moves and every user stays in the same room. The grouping server 
immediately starts the device grouping when all devices are 
connected. The parameters for the static simulation are shown in 
\prettyref{tab:parameters-device-grouping}. We run the device 
grouping using random light patterns with different length $\in$ 
\{2, 4, 6, 8, 10\} and from at least two users up to ten users.

\textbf{Simulation results} Through 10-fold cross-validation, 
highlighted in bold in \prettyref{tab:results-grouping}, we 
identify the best working device grouping technique in relation 
to a fast reasoning and a reasonable overall result, i.e., the 
average over accuracy, precision, recall, and F1-score. The 
device localization using Bluetooth and Wi-Fi features works 
worst whereas ML-based device grouping performs similar or 
slightly better than the signal similarity metrics such as 
Spearman and Pearson. Moreover, 
\prettyref{fig:runtime-static-coupling} presents 
the runtime of each method for device grouping sorted in 
ascending order. To perform device grouping, it takes around 
1.41\,s to receive data (83.93\,\% of total time) compared to 
the actual device grouping with 0.27\,s (16.07\,\% of the total 
time).

For a thorough evaluation, we further analyze the performance of 
light patterns with different lengths for device grouping. The 
light pattern with four random on and off periods works best 
compared to a decrease of 9\,\% using the worst light pattern 
with ten random periods. The performance of light patterns with 
different lengths sorted by descending total result in brackets, 
i.e., average over accuracy, precision, recall, and F1-score: 4 
(0.97), 2 (0.95), 6 (0.91), 8 (0.9), and 10 (0.88). Furthermore, 
we evaluate the performance using different number of grouping 
users. With six users we reach the highest total result of 0.97 
because with less users the grouping signals miss crucial 
patterns and with more users the noise in the grouping signals 
grows which leads to a higher error rate for device 
associations. In detail, we show the number of grouping users 
with descending total result in brackets: 6 (0.97), 4 (0.94), 3 
(0.93), 8 (0.93), 5 (0.92), 9 (0.92), 10 (0.92), 7 (0.92), and 2 
(0.87).

\subsection{Dynamic Device-to-Area Simulation of Device 
Grouping}

\textbf{Simulation settings} In contrast to the static 
simulation, in the dynamic device-to-area simulation the users 
are moving between different rooms receiving varying light 
patterns. Our modeled simulation environment for device 
associations is shown in \prettyref{fig:simulation-environment}. 
The rooms are positioned in a rectangular arrangement with an 
inter room distance of 3\,m and intra room distance of 2\,m. We 
compute the distances among all room combinations. Using the 
duration of one simulation iteration of 20\,min, we calculate a 
random path between the rooms for each user. Thereby, we 
distribute the random time as duration of stay over different 
rooms using a multinomial distribution. As a result, the user's 
random path is a list of tuples with duration of stay and room, 
e.g., user A has the random path [(1, 120), (3, 300), ...]. This 
means that the start position is in room 1 and after 120\,s the 
user moves to room 3 and stays there for 5\,min, and so forth. 
Hence, we randomly create user groups for each room at a 
specific time and for each movement between rooms every user 
chooses a random movement speed in the range of 1.25 to 
1.53\,m/s (4.5--5.5\,km/h) \cite{Carey.2005}. If the users are 
in motion between two rooms they are in the corridor and not 
associated with any room. For device grouping, each room acts 
independently of other rooms and is linked with a unique 
location-dependent environment data containing Wi-Fi and 
Bluetooth scans, and light patterns. The overview of parameters 
for dynamic device-to-area simulation in 
\prettyref{tab:parameters-device-grouping} covers grouping 
frequency, number of users, and number of rooms.

\textbf{Simulation results} \prettyref{tab:results-grouping} 
shows the best working device grouping (highlighted in bold) 
using 10-fold cross validation in terms of a fast runtime and a 
reasonable overall result, meaning the average over accuracy, 
precision, recall, and F1-score. Compared to the static 
device-to-device simulation, the device grouping based on 
similarity metrics works slightly better than ML-based device 
grouping. Further, the device localization using Wi-Fi and 
Bluetooth features reaches a similar result. The runtime of each 
method for device grouping is shown in 
\prettyref{fig:runtime-dynamic-coupling} with ascending runtime 
from 0.26\,s to 1.58\,s. The median time is around 0.43\,s 
(71.67\,\% of the total time) to receive data for device 
grouping whereas the device grouping itself lasts 0.17\,s 
(28.33\,\% of the total time). Besides that, the frequency of 
device grouping with 20\,s works best, the accuracy of device 
grouping decreases by 16\,\% with 30\,s and with a 10\,s 
frequency the accuracy decreases another 8\,\%. Furthermore, we 
analyze the performance of device grouping with a varying number 
of rooms, sorted after decreasing overall result in brackets: 1 
(0.99), 2 (0.96), 3 (0.92), 5 (0.9), 6 (0.89), 4 (0.87), 8 
(0.84), 7 (0.82), 9 (0.79), and 10 (0.76). The device grouping 
works 
best with less rooms because the more rooms the higher the risk 
that the user miss the up-to-date light pattern of the 
designated room due to movement between rooms.

Summarizing, \prettyref{tab:results-grouping} presents two 
different best working classifiers and features for device 
grouping depending on the use case either with static or moving 
users. In both cases, we have an equal distribution between 
machine learning based and similarity based device grouping. 
Moreover, similar grouping techniques perform best, mainly the 
feature types are changing in varying conditions. Nevertheless, 
we favor the approach for device grouping in the scenario with 
multiple moving users because it is more realistic in practice.

\section{Practical Extension of DevLoc}
DevLoc provides the basis to connect devices for data sharing 
among users. As a practical extension to enhance data privacy 
and ease the setup of data sharing, we analyze logs of device 
associations in terms of grouping patterns, e.g., time and 
frequency, to detect semantic device groups like personal and 
stranger's devices. On this basis, we are able to create 
automated data sharing policies, such as sharing sensitive data 
only among personal devices.



\subsection{Artificial Log of Device Associations}
In practice, DevLoc records device associations for further 
analysis to find semantic device groups. For evaluation we 
generate an artificial log of device associations to be able to 
analyze the device groups across different testbeds instead of 
using real-world data sets \cite{copelabs-usense-20170127, 
upb-hyccups-20161017} of social networks. To generate the log of 
device associations, we define a calendar for the simulation 
time including days $\in$ \{all, holiday, weekend, workday\} and 
time slots structuring the hours of the day: all $\in [0, 24]$, 
night morning $\in [0, 5]$, morning $\in [5, 10]$, forenoon $\in 
[10, 12]$, noon $\in [12, 14]$, afternoon $\in [14, 17]$, 
evening $\in [17, 21]$, and evening night $\in [21, 24]$. On 
this basis, we specify three different device groups $\in$ 
\{personal, family \& friends, well-known \& stranger\} with 
corresponding time encounter rules. For personal and family \& 
friends devices: morning[workdays], evening[workdays], 
noon[all], afternoon[all], evening[all] in contrast to 
well-known \& stranger devices where all encounter times are 
allowed: all[all], i.e., entirely random device encounters. We 
determine 
the possible device encounters based on the corresponding rule 
of the device group and then randomly distribute the device 
encounter time over the simulation duration, in our case one 
year (365 days). Finally, we clean the generated log by removing 
log entries with single and duplicated devices. 


\begin{table*}
	\caption{Best working classifiers and features to predict 
		semantic device groups}
	\label{tab:device-log-ml}
	\centering
	\begin{tabular}{llllcccccc}
		\toprule
		Device group & Testbed & Classifier & Feature type & AUC 
		& Cold start (days) & Accuracy & Precision & Recall & 
		F1-score
		\\
		\midrule
		\multirow{3}{*}{Personal} & Dense & Naive bayes & 
		Contact frequency per week - sum & .99 & 23 & 
		.95 & .92 & .93 & .92
		\\
		\cmidrule{2-10}
		& Medium & Extra trees & Grouping time per week - mean & 
		.97 & 17 & .93 & .87 & .92 & .89
		\\
		\cmidrule{2-10}
		& Sparse & Ada boost & Contact frequency per week & 
		.98 & 56 & .96 & .94 & .93 & .94
		\\
		\midrule
		\multirow{3}{*}{\shortstack[l]{Family \\ \& Friends}} & 
		Dense & Naive bayes & Contact frequency per week - sum & 
		.99 & 17 & .94 & .92 & .92 & .92
		\\
		\cmidrule{2-10}
		& Medium & Gradient boosting & Grouping time per week - 
		std & .85 & 72 & .89 & .84 & .81 & .83
		\\
		\cmidrule{2-10}
		& Sparse & Ada boost & Contact frequency per week & .98 
		& 55 & .96 & .94 & .95 & .95
		\\
		\midrule
		\multirow{3}{*}{\shortstack[l]{Well-known \\ \& 
				Stranger}} & Dense & Ada boost & Grouping time 
				per week 
		- sum & .98 & 16 & .97 & .97 & .93 & .95
		\\
		\cmidrule{2-10}
		& Medium & Ada boost & Contact frequency per week & .99 
		& 22 & .97 & .95 & .95 & .95
		\\
		\cmidrule{2-10}
		& Sparse & Ada Boost & Contact frequency per week & 
		.99 & 45 & .98 & .97 & .96 & .96
		\\
		\bottomrule
	\end{tabular}
\end{table*}

\subsection{Semantic Log Analysis of Device Associations}
\textbf{Features for Log Analysis} We create 43 different 
feature sets of three or ten dimensional features using the 
following information:
\begin{itemize}
	\item how much time the devices are associated together per 
	event, per day, and per week,
	\item how often the devices are associated together per day 
	and per week, and 
	\item the time ratio between association time and entire 
	time frame per day and per week.
\end{itemize}

In addition, we calculate statistical measures like average and 
standard deviation
of multiple combined features per event and per week. We 
use the statistical measures all together in a ten dimensional 
feature vector or we treat them individually as three 
dimensional feature vectors.

\textbf{Testbeds for Log Analysis} To simulate different test 
environments and compare the results, we introduce three 
environments with varying numbers of devices per device group 
and different grouping times, could be also more groups. The 
sparse environment includes three devices per device group and a 
grouping time in the range [10, 60]\,min, the medium environment 
has six devices per group and a grouping time of [20, 120]\,min, 
and the dense environment uses a grouping time of [30, 180]\,min 
and nine devices per device group.

\begin{figure}
	\centering
	\begin{subfigure}[b]{\linewidth}
		\centering
		\begin{subfigure}[b]{0.325\linewidth}
			\caption*{Testbed: dense}
			\includegraphics[width=\linewidth]{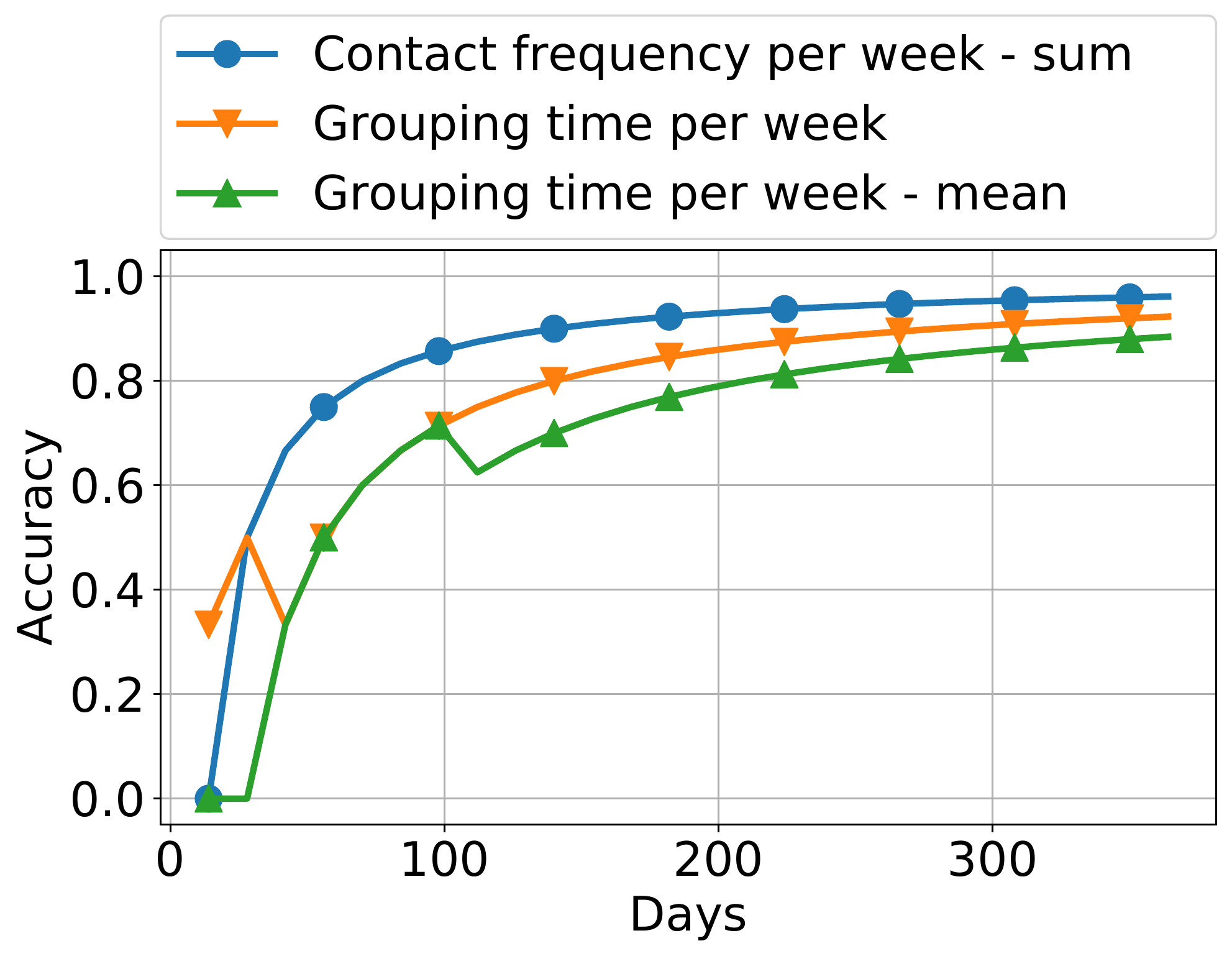}
		\end{subfigure}
		\hfill
		\begin{subfigure}[b]{0.325\linewidth}
			\caption*{Testbed: medium}
			\includegraphics[width=\linewidth]{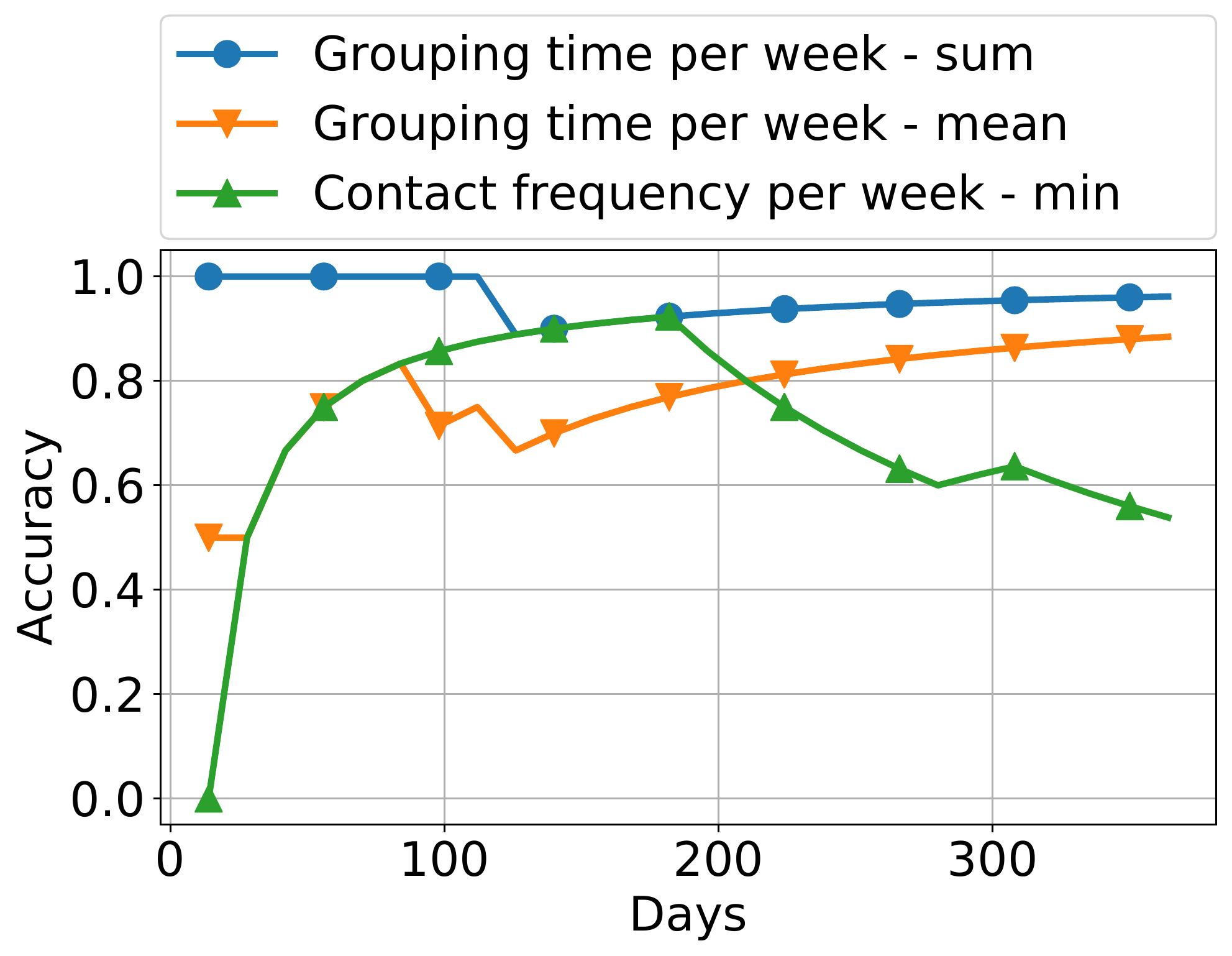}
		\end{subfigure}
		\hfill
		\begin{subfigure}[b]{0.325\linewidth}
			\caption*{Testbed: sparse}
			\includegraphics[width=\linewidth]{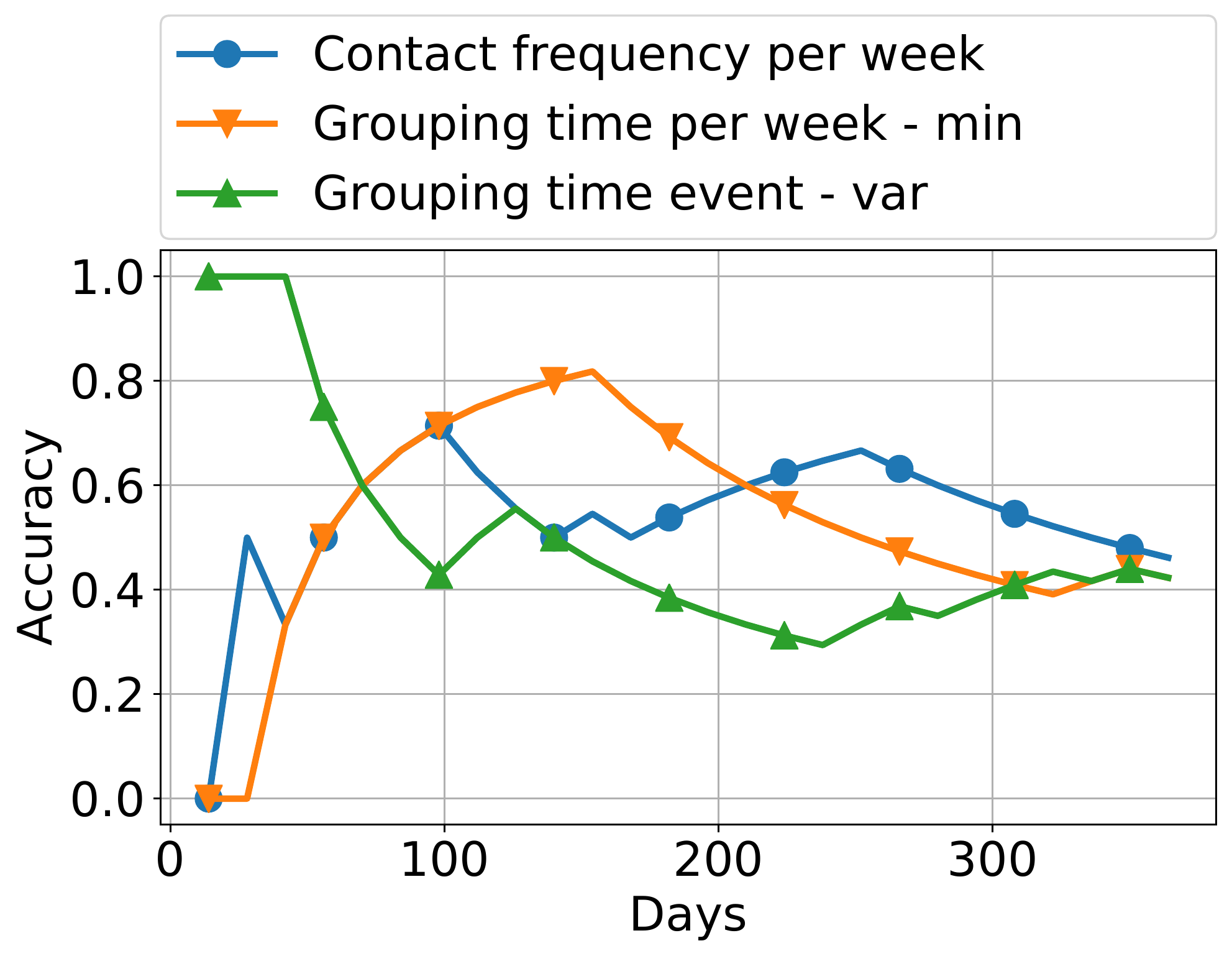}
		\end{subfigure}
		\caption{Detection accuracy of the three best performing 
		features to cluster devices into three single device 
		groups}
	\end{subfigure}
	\\
	\vspace{2mm}
	\begin{subfigure}[b]{\linewidth}
		\centering
		\begin{subfigure}[b]{0.325\linewidth}
			\caption*{Testbed: dense}
			\includegraphics[width=\linewidth]{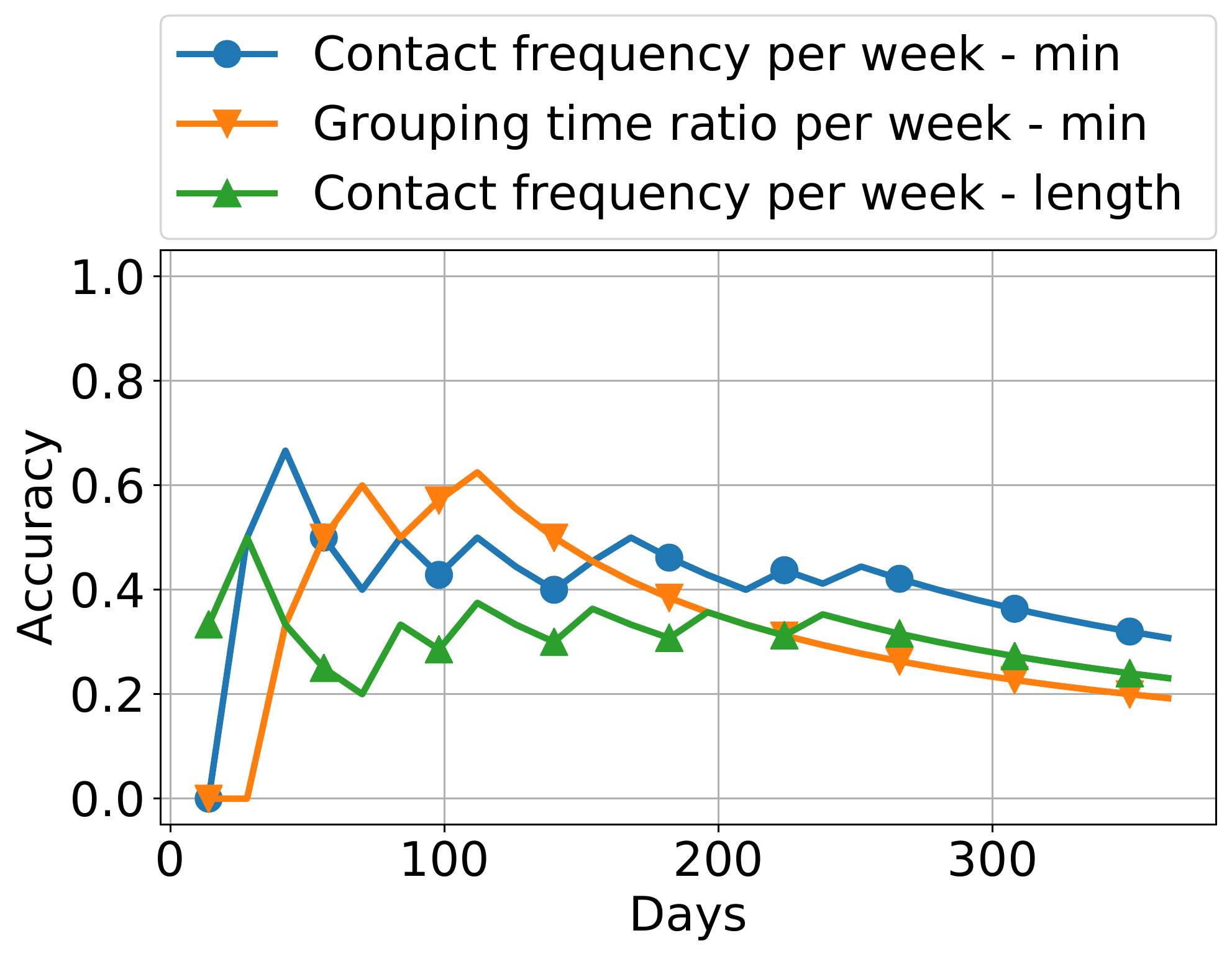}
		\end{subfigure}
		\hfill
		\begin{subfigure}[b]{0.325\linewidth}
			\caption*{Testbed: medium}
			\includegraphics[width=\linewidth]{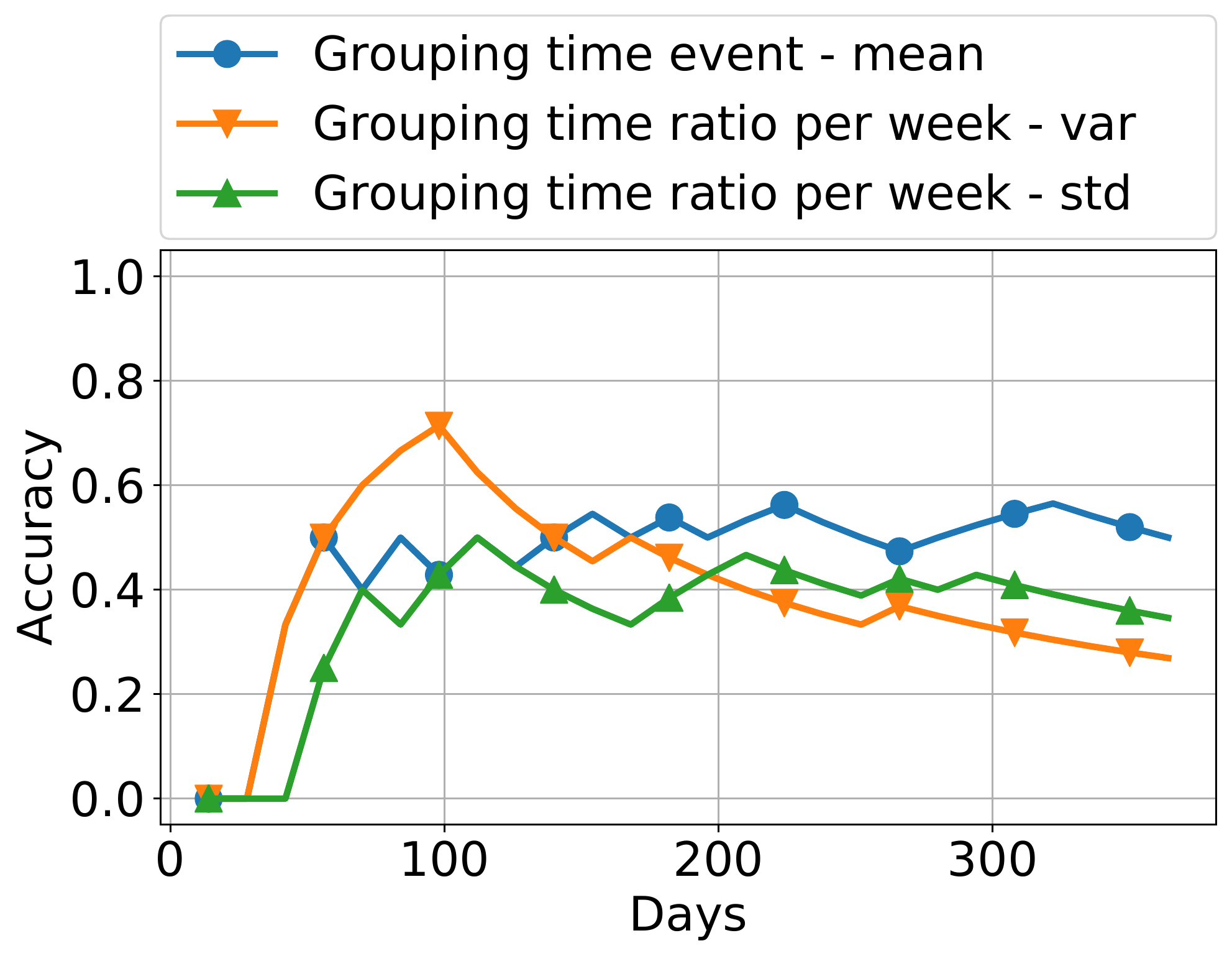}
		\end{subfigure}
		\hfill
		\begin{subfigure}[b]{0.325\linewidth}
			\caption*{Testbed: sparse}
			\includegraphics[width=\linewidth]{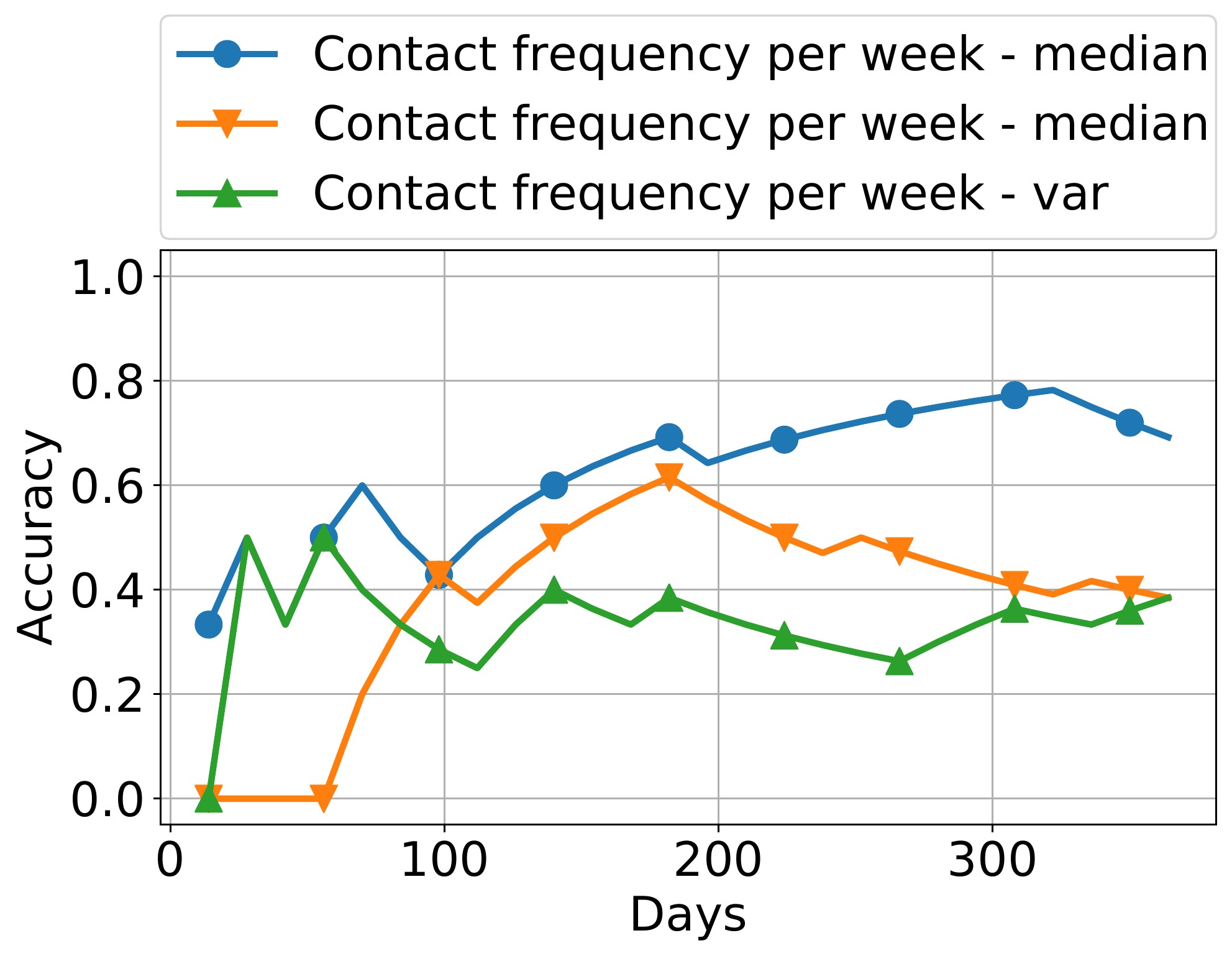}
		\end{subfigure}
		\caption{Detection accuracy of the three best performing 
		features to cluster devices into seven mixtures of 
		device groups}
	\end{subfigure}
	\caption{Detection granularity of different device groups 
	over time: personal, family \& friends, and well-known \& 
	stranger across testbeds with varying numbers of devices: 
	dense, medium, and sparse.}
	\label{fig:device-log-clustering}
\end{figure}

\textbf{Detection granularity of device groups over time} 
\prettyref{fig:device-log-clustering} shows the clustering 
accuracy over time for each test environment with single or 
mixtures of device groups. We sort each cluster estimation 
after the mean accuracy in descending order and select the 
three best performing clustering methods. Our results show that 
in the dense and medium testbeds, we are able to reliably 
identify the three single device groups, in the sparse testbed 
the input data is not sufficient to estimate the device groups 
over time. The same holds true to detect the seven mixtures of 
device groups, over all testbeds it is impossible to reliably 
recognize the device groups. Our further analysis is restricted 
to three single device groups, if we can predict these device 
groups, we can also define data policies for a mixture of them.

In detail, at each time step $t$ we calculate the accuracy 
based on the estimated number of clusters via different 
clustering methods and the true number of clusters. Thereby, we 
highlight the temporal behavior to detect the number of device 
groups based on the log of device associations. Our log includes 
either three single device groups: personal, family \& friends, 
well-known \& stranger or seven mixtures of device groups, 
e.g., personal + family \& friends. We use the following 
clustering methods to 
the range between 2 to 9 clusters:
\begin{itemize}
	\item K-Means with elbow method using the total within sum 
	of squares to measure the cluster compactness
	\item K-Means using silhouette score
	\item Hierarchical clustering using silhouette score
	\item Gaussian mixture using Bayesian Information Criterion 
	\item Gaussian mixture using Akaike Information Criterion 
	\item X-Means using K-Means++ for initial cluster centers
\end{itemize}

\textbf{Find the best working features and classifiers to 
predict device groups} \prettyref{tab:device-log-ml} presents 
for each device class and testbed the best working classifier 
and feature type including the average area under the curve 
(AUC), cold start in days, accuracy, precision, recall, and 
F1-score. We model the problem to identify the best working 
classifiers and features to predict device groups as multi-class 
classification with the following device classes: 0 -- personal, 
1 -- family \& friends, and 2 -- well-known \& stranger.
In total, we use 43 feature sets with a series of different 
classifiers via 10-fold cross validation. The classifiers 
include extra trees, gradient boosting, SVM, random forest, 
naive Bayes, and Ada boost. The cold start is the success 
criterion meaning after which time we are able to reliably 
predict the device class, the earlier the better. The cold start 
is defined by a threshold of 80\,\%, from this point in time 
(day), all result metrics including accuracy, precision, recall, 
and F1-score are above this threshold. Furthermore, we consider 
only prediction results with a cold start in the first quarter 
of the overall timeline of one year.





\begin{figure}
	\centering
	\begin{subfigure}[b]{0.24\linewidth}		
		\includegraphics[width=\linewidth]{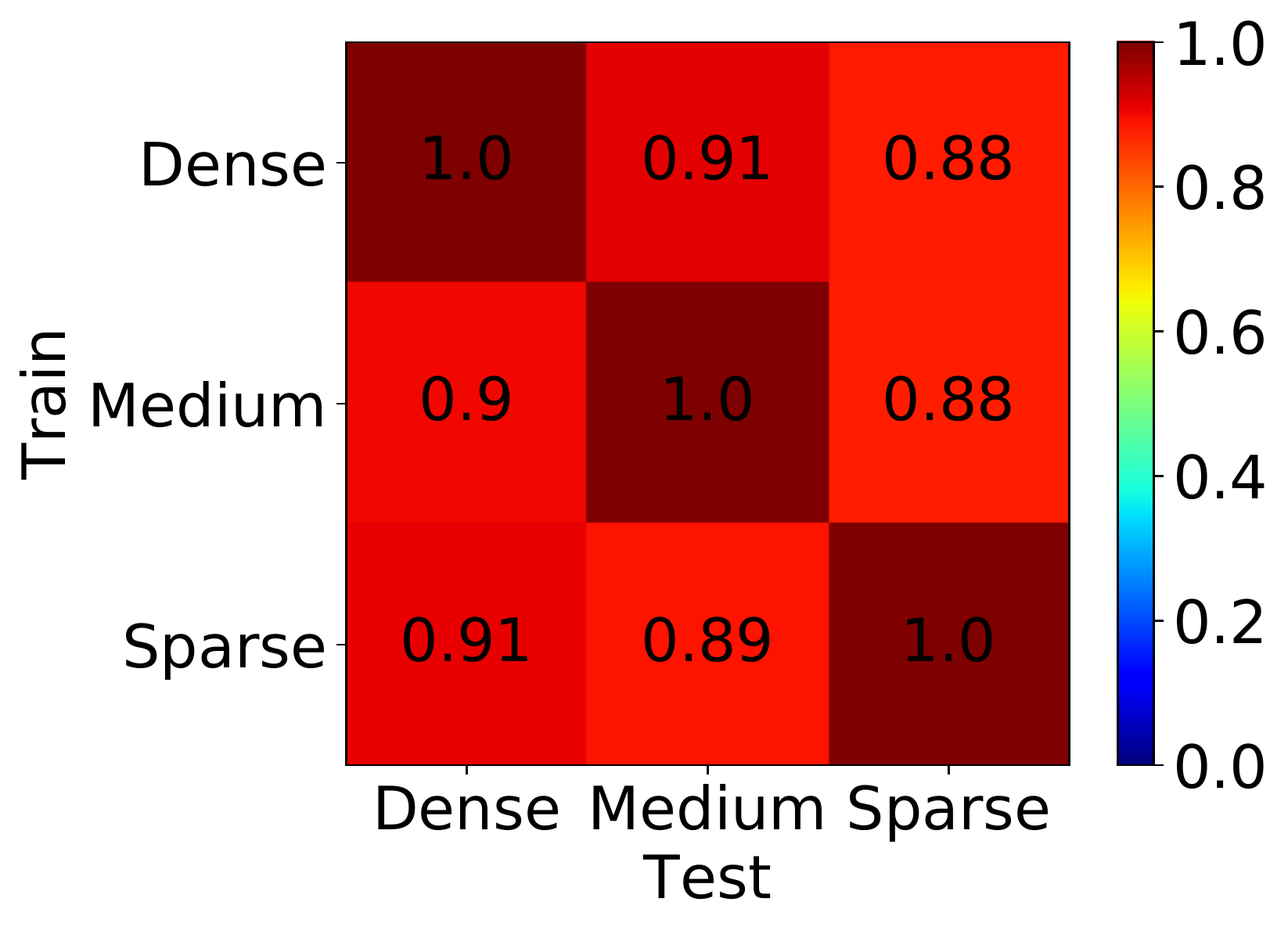}
		\caption{Accuracy}
		\label{fig:coupling-log-across-testbeds-accuracy}
	\end{subfigure}
	\hfill
	\begin{subfigure}[b]{0.24\linewidth}
		\includegraphics[width=\linewidth]{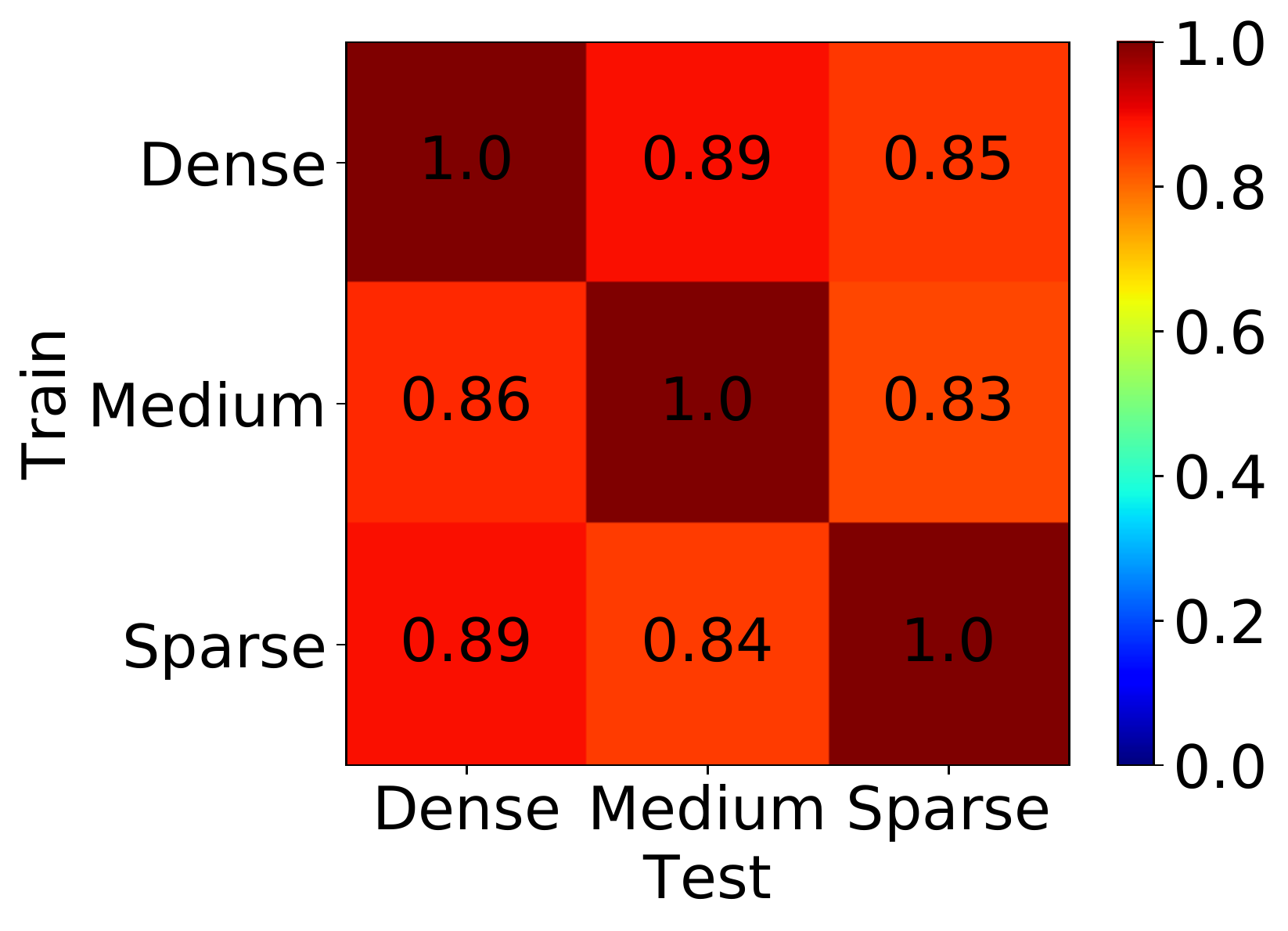}
		\caption{Precision}
		\label{fig:coupling-log-across-testbeds-precision}
	\end{subfigure}
	\hfill
	\begin{subfigure}[b]{0.24\linewidth}		
		\includegraphics[width=\linewidth]{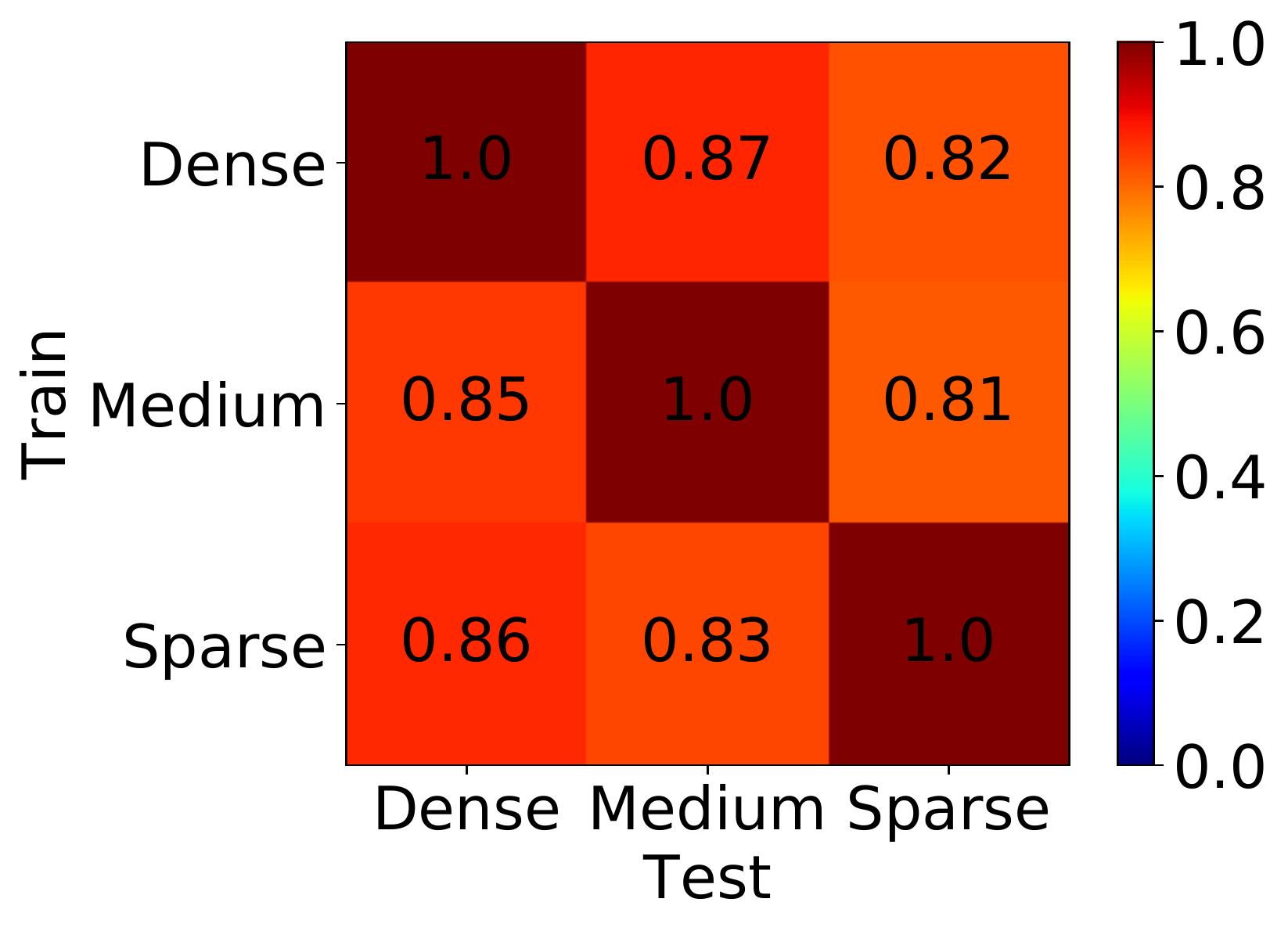}
		\caption{Recall}
		\label{fig:coupling-log-across-testbeds-recall}
	\end{subfigure}
	\hfill
	\begin{subfigure}[b]{0.24\linewidth}	
		\includegraphics[width=\linewidth]{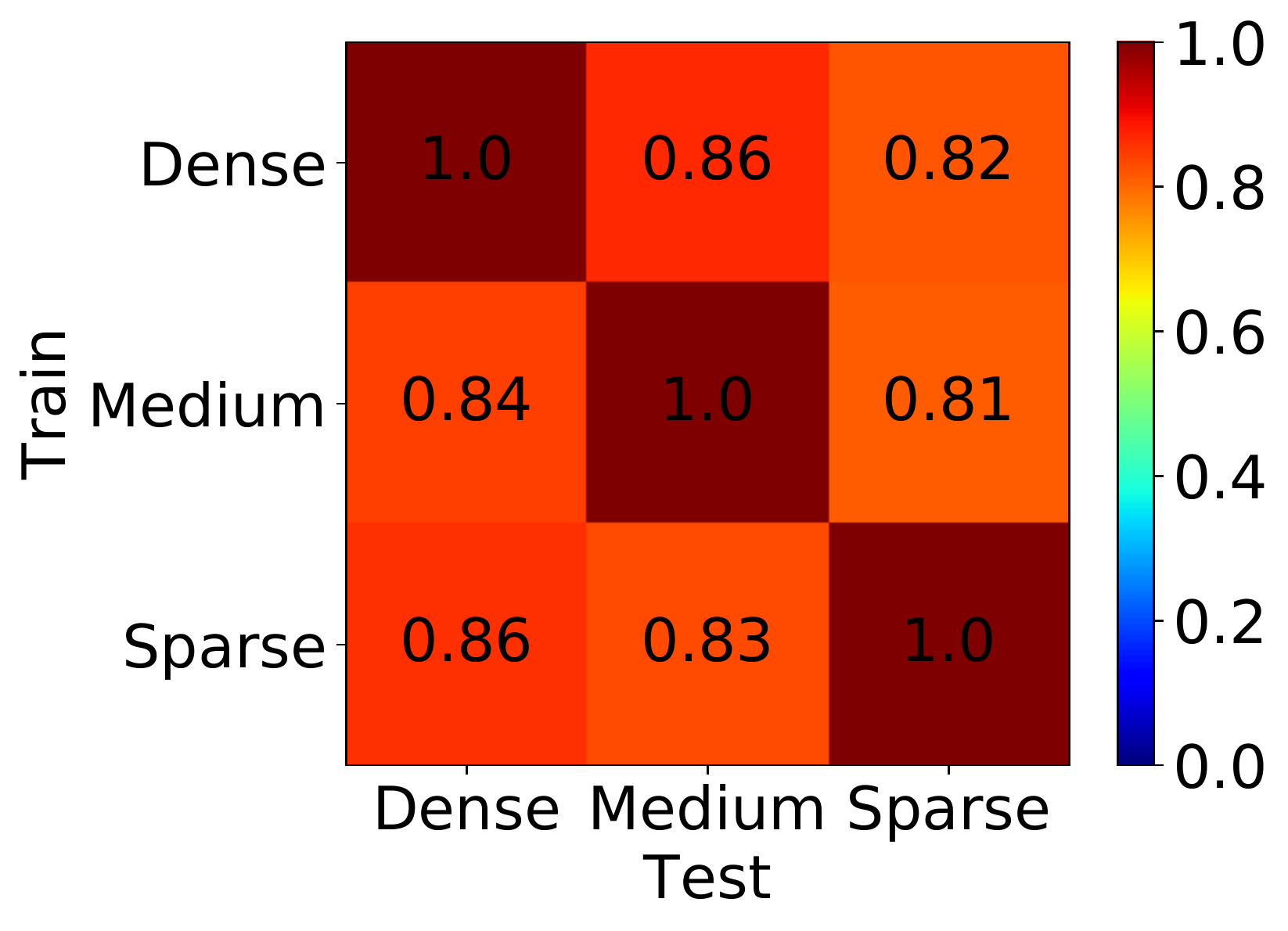}
		\caption{F1-score}
		\label{fig:coupling-log-across-testbeds-f1}
	\end{subfigure}
	\caption{Average prediction result of the best performing 
	classifiers and features to infer semantic device groups 
	across testbeds with different number of devices}
\end{figure}

\textbf{Stability to predict devices groups over time and across 
testbeds} 
\prettyref{tab:coupling-log-across-testbeds-classifiers-features}
shows the most common combinations of classifier and feature per 
train and test environment among best performing prediction 
methods for semantic device groups. To be specific, we identify 
which combination of classifier and feature type works best over 
several test environments with varying numbers of devices and 
different device encounter times. Via 10-fold cross-validation, 
we train and test the classifier for each combination of 
\{dense, medium, sparse\} testbed and time of testing. 
\crefrange{fig:coupling-log-across-testbeds-accuracy}{fig:coupling-log-across-testbeds-f1}
illustrate the average prediction result of the best performing 
classifiers and features, i.e., highest average detection 
accuracy over all device classes: personal, family \& friends, 
and well-known \& stranger to predict semantic device groups.

\begin{table*}
	\centering
	\caption{Most common combination of classifier and feature 
	among best prediction methods for semantic device groups}
	\label{tab:coupling-log-across-testbeds-classifiers-features}
	\begin{tabular}{cccc}
		\toprule
		\diagbox{Train}{Test} & Dense & Medium & Sparse
		\\
		\cmidrule{2-4}
		Dense & Ada Boost + Contact frequency per day & Naive 
		Bayes + Grouping time ratio per week & Naive Bayes + 
		Grouping time ratio per week
		\\
		\midrule
		Medium & SVM + Grouping time per week - min & Ada Boost 
		+ Contact frequency per day & SVM + Grouping time per 
		week - mean
		\\
		\midrule
		Sparse & SVM + Grouping time per week & SVM + Grouping 
		time per week & Ada Boost + Contact frequency per day
		\\
		\bottomrule
	\end{tabular}
\end{table*}

We encounter a potential privacy leakage by storing and 
analyzing device associations to infer semantic device groups. 
Countermeasures can be the anonymization of device activity 
logs, store the activity logs for a short period of time, and 
blockchain-secured device logging, or perform analysis in home 
only locally on the access point.

We conclude the evaluation of DevLoc to enable seamless device 
grouping based on visible light signaling. First, we analyze the 
characteristics of the VLC physical channel and afterwards we 
perform a thorough parameter estimation including pre-selection 
of distance and correlation measures for light patterns and 
feature selection for ML-based device association. On this 
basis, we run two device grouping simulations with a single room 
and static users and several rooms with moving users to find for 
each case the best working device grouping method. Besides that, 
we use the log of device associations to infer semantic device 
groups like personal, family or stranger's devices to support 
data sharing policies, with whom sharing which data.
For future extension, we plan to calculate a trust score 
\cite{Chatzopoulos.2016} based on the log of device interactions 
from DevLoc to further enrich the semantic meaning of devices 
regarding allowed data sharing among devices.

\section{Discussion}
Due to the configuration framework of DevLoc based on visible 
light signaling integrated in surrounding lighting, we 
can support fine-granular device associations per room or 
region. In this way, we can realize our previously defined use 
cases for end users and IoT applications. By using the 
similarity of distance-limited light patterns, we are able to 
help Alice asking for the best way on the subway or Carol to 
record who is present at here meetings. Combined with LocalVLC 
\cite{Haus.2019b}, to incorporate our Morse-code inspired 
modulation scheme 
that can operate on off-the-shelf LEDs with low energy overhead, 
we are able to transmit data via light, e.g., location 
identifier, and not only light patterns. Thereby, we can support 
IoT applications, such as merging and filtering of 
location-tagged data from IoT boards and location-based access 
policies for consumer smart home platforms.




To enhance the user's privacy for DevLoc, we will use private 
proximity testing at the light bulb during device grouping. 
Thereby, as part of the secure multi-party computation (SMC) 
problem, multiple parties are able to compute whether they are 
nearby without learning each other's inputs.
Homomorphic encryption \cite{Gentry.2009} and garbled circuit 
\cite{Yao.1982} are two main techniques to solve the SMC 
problem, at which homomorphic encryption is more efficient 
compared to garbled circuit \cite{Hallgren.2015}. Usually 
homomorphic encryption is applied to a small amount of 
data, e.g., latitude and longitude, to compute the distance 
between two points of interest. In the appendix we analyze the 
runtime of fully homomorphic encryption for time-series data 
such as light patterns and we pinpoint the need of new 
cryptographic primitives for practical use.

Moreover, we plan to enhance DevLoc to be more resilient 
against adversaries performing relay attacks to trick our device 
grouping to include distant clients into the device group by 
relaying location-dependent light patterns from the intended 
space to remote clients. Besides that, we have to consider 
mitigation strategies against other attacks like spoofing the 
area's light reference signal that clients use for association
or periodically starting the device grouping every few seconds 
may introduce a frequent opportunity for potential attackers to 
identify themselves as part of the group.

\section{Conclusion}
DevLoc is a ready-to-use system solution that provides a 
seamless device grouping based on visible light signaling for 
data sharing. Our customized light bulb 
transmissions allows clients to detect cycles in the light 
patterns for device grouping.
We perform a thorough evaluation of DevLoc via two different 
simulations with a single room and static users and multiple 
rooms with moving users. Thereby, we analyze the performance of 
several device grouping methods: signal similarity based on 
distance and correlation metrics, ML-based signal similarity, 
and as baseline the device localization using Wi-Fi and 
Bluetooth traces. Finally, we take advantage of the device 
grouping log to infer semantic device groups like personal, 
family or stranger's devices to enhance the data privacy, with 
whom sharing which data.

\balance
\bibliographystyle{IEEEtran}
\bibliography{references/main,references/external,references/llcm-stateofart}

\pagebreak
\appendix

\textbf{Practicality of Fully Homomorphic Encryption for 
Private Proximity Testing}
To protect the user's privacy during device grouping at the 
light bulb, we apply private proximity testing as an instance of 
the secure multi-party computation (SMC) problem where multiple 
parties compute whether they are nearby within a specific 
distance threshold without learning each other's inputs. This 
protects the location data, e.g., light patterns, against a wide 
range of attacks, because it reveals no sensitive information to 
anyone. Homomorphic encryption \cite{Gentry.2009} and garbled 
circuit \cite{Yao.1982} are two main techniques to solve the SMC 
problem, at which homomorphic encryption is more efficient 
compared to garbled circuit \cite{Hallgren.2015}.

Our system model for proximity detection consists of a trusted 
party, a service provider, mobile users, and static IoT devices. 
The service provider is usually considered untrusted and should 
not learn the proximity test results. Our custom light bulb acts 
as trusted party and service provider, e.g., data sharing, among 
the nearby mobile devices. Usually homomorphic encryption is 
applied to a small amount of data to compute the distance 
between two points of interest, e.g., latitude and longitude of 
attractions like for LBS. In contrast, we analyze the runtime of 
fully homomorphic encryption for time-series data such as light 
patterns whether it is practically usable.


\textbf{Testbeds} To evaluate the performance of homomorphic 
encryption, we use the libraries: HElib \cite{helib} and SEAL 
\cite{sealcrypto} to compute the euclidean and cosine distance 
as similarity measure between two light patterns on three 
different platforms: server, NUC, and IoT board as described in 
\prettyref{tab:hardware-testbeds}.

\begin{table}[h]
	\centering
	\caption{Testbed's hardware specifications for runtime 
	analysis of fully homomorphic encryption}
	\label{tab:hardware-testbeds}
	\begin{tabular}{L{2.6cm}ll}
		\toprule
		System &
		CPU &
		RAM
		\\
		\midrule
		Server & 40x Intel Xeon E5-2630 2.2\,GHz & 768\,GB 
		\\
		\midrule
		Virtual machine (VM) & 4x Intel Xeon E5-2630 2.2\,GHz 
		& 32\,GB
		\\
		\midrule
		Next unit of computing PC (NUC) & 4x Intel Core i5-6260U 
		1.8\,GHz & 16\,GB
		\\
		\midrule
		IoT board & 1x ARM AM3358 1\,GHz & 0.5\,GB
		\\
		\bottomrule
	\end{tabular}
\end{table}

\begin{figure}[h]
	\centering
	\includegraphics[width=0.85\linewidth]{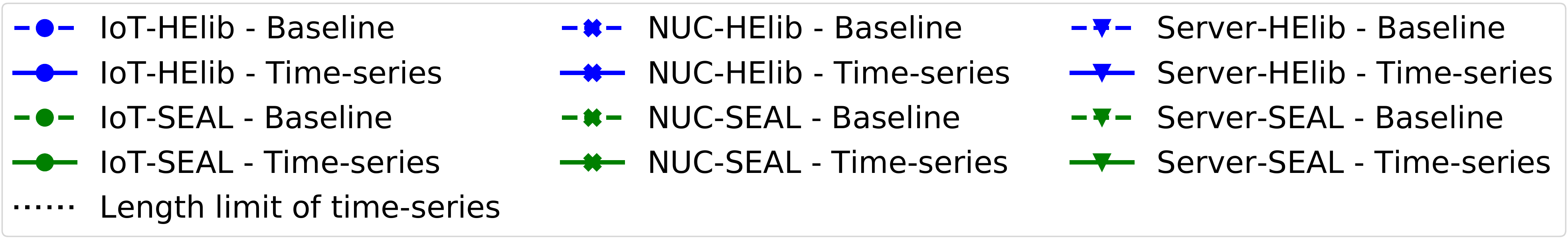}
	\\
	\begin{subfigure}[b]{0.32\linewidth}
		\includegraphics[width=\linewidth]{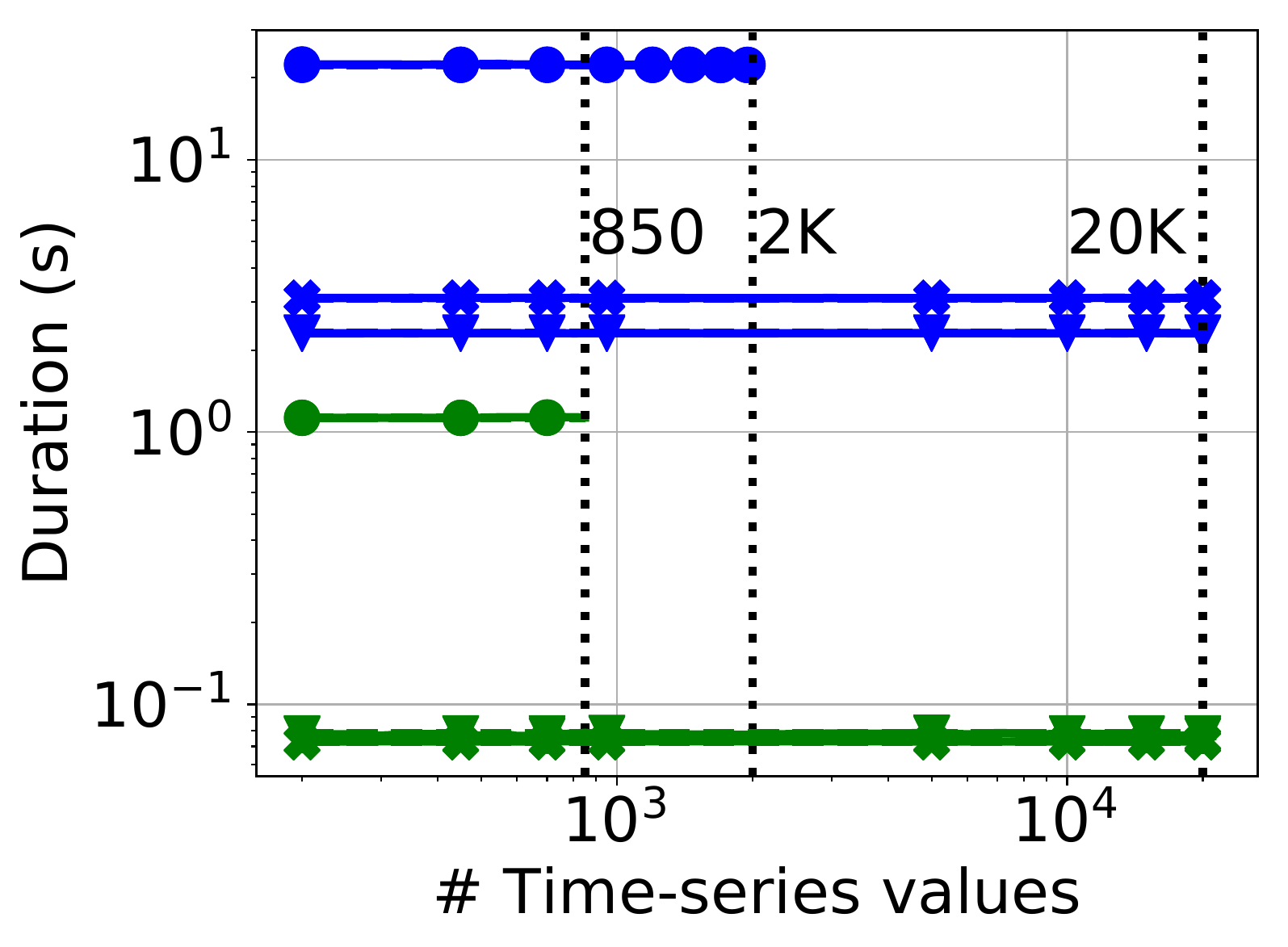}
		\caption{Initialization}
		\label{fig:he-performance-init}
	\end{subfigure}
	\hfill
	\begin{subfigure}[b]{0.32\linewidth}
		\includegraphics[width=\linewidth]{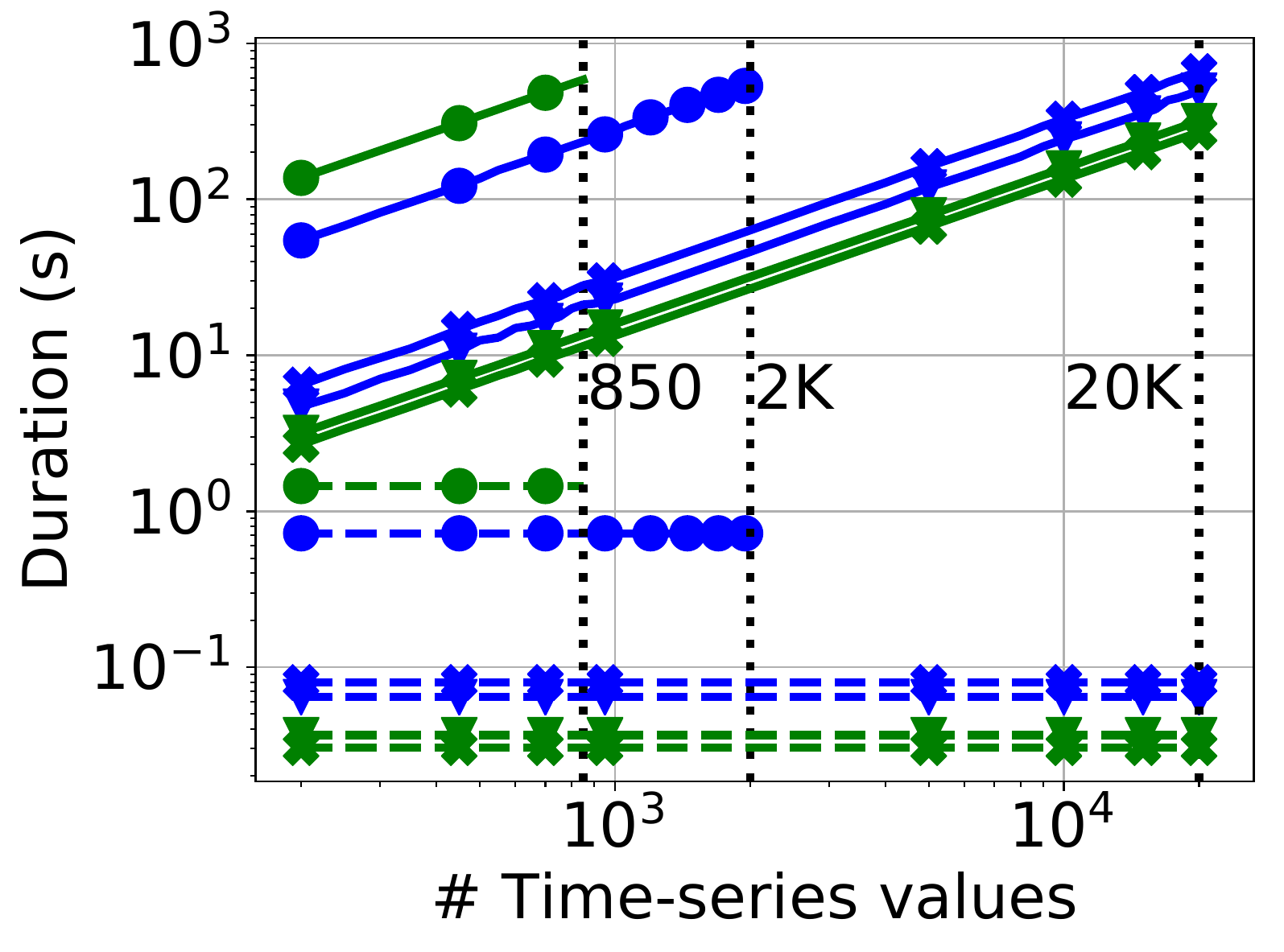}
		\caption{Euclidean distance}
		\label{fig:he-performance-euclidean}
	\end{subfigure}
	\hfill
	\begin{subfigure}[b]{0.32\linewidth}
		\includegraphics[width=\linewidth]{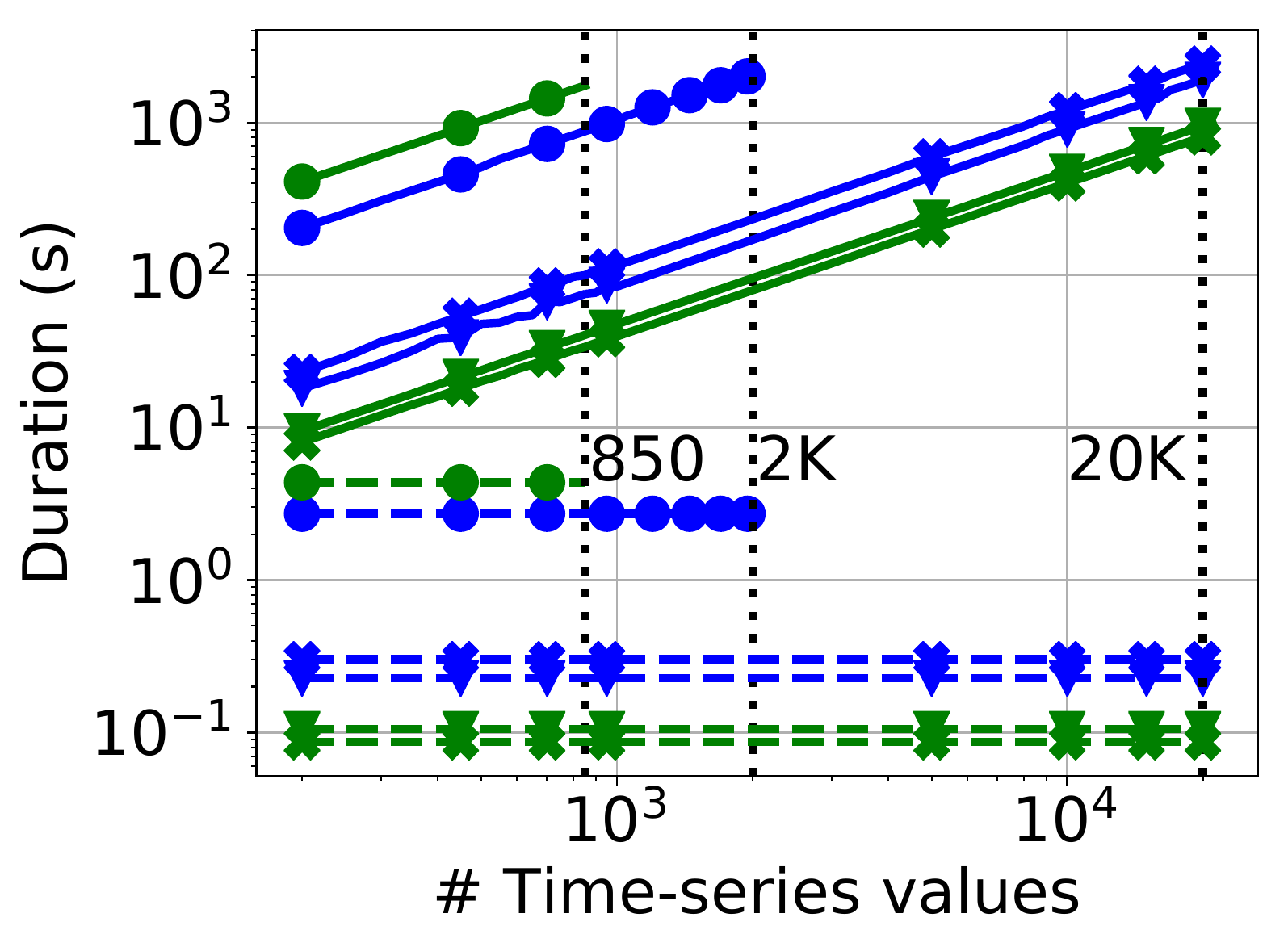}
		\caption{Cosine distance}
		\label{fig:he-performance-cosine}
	\end{subfigure}
	\caption{Performance of fully homomorphic encryption over 
	three different testbeds: server, NUC, and IoT board using 
	two libraries: HElib and SEAL. The baseline refers to the 
	distance calculation using only two-dimensional positions 
	compared to time-series with hundreds of values.}
\end{figure}

\textbf{Performance results} 
\crefrange{fig:he-performance-init}{fig:he-performance-cosine} 
show the duration to initialize two different time-series each 
with a size in the range of [200, 21k] and compute the euclidean 
and cosine distance for proximity detection. We encounter 
data limits due to restricted system's memory: on the IoT board 
the SEAL library achieves a maximum time-series length of 
850 values compared to HElib with a size of 2k, and on the 
NUC only HElib reaches a data limit of 20k values. We analyze 
the library performance per test platform over all 
operations $\in$ \{initialization, euclidean, cosine\} and 
time-series values. The HElib library works fastest at the 
server, the NUC is about 33\,\% slower, and IoT board is 11.7 
times slower compared to the NUC. In comparison, the SEAL 
library performs best at the server, the NUC is 19\,\% slower, 
and the IoT board takes 51 times longer.

We compute the runtime per operation over all test 
platforms and time-series values. The initialization of 
time-series vectors takes on average 0.43\,s using SEAL 
and 9\,s with HElib, hence HElib is around 11.67 times slower 
per vector element. For the euclidean distance, the 
HElib library requires between [36, 569]\,s and the SEAL 
library achieves [56, 397]\,s, the relative runtime per 
vector element of HElib with 0.11\,s is 2.16 times faster 
compared to SEAL (0.24\,s). For the cosine distance, the 
HElib library takes on average between [135, 2137]\,s 
and using the SEAL library lasts [166, 1187]\,s, the 
relative runtime per vector element of HElib with 0.42\,s 
is 1.72 times faster compared to SEAL with 0.72\,s.

We compare the runtime performance of the baseline using 
two-dimensional positions like for LBS with our generated light 
patterns ranging between 1762 to 3717 raw voltage signals. We 
compute the average runtime for each homomorphic encryption 
library over all test platforms and operations at which the 
light pattern contains on average 1417 voltage values. The 
HELib library is slower with a runtime of 0.72\,s for the 
baseline and the light pattern takes around 63.39\,s, with 
a normalized delta of 90.99\,\% over the time-series 
length. The faster SEAL library achieves a runtime of 
0.09\,s for the baseline and 31.8\,s for the larger 
light signal which results in a normalized delta of 63.34\,\%.

In a nutshell, our aim is to enhance user's privacy by applying 
homomorphic encryption to secure the time-series data processing 
of our device grouping. The runtime analysis 
motivates the need of new cryptographic primitives for efficient 
time-series data analysis. The up-to-date homomorphic encryption 
is too slow to be usable in practice with a runtime of about 
30\,s per distance computation whereas we require a maximum 
runtime of 0.5\,s per calculation.

\end{document}